\documentclass[fleqn,usenatbib]{mnras}

\usepackage{newtxtext,newtxmath}

\usepackage[T1]{fontenc}

\DeclareRobustCommand{\VAN}[3]{#2}
\let\VANthebibliography\thebibliography
\def\thebibliography{\DeclareRobustCommand{\VAN}[3]{##3}\VANthebibliography}

\usepackage{graphicx}	
\usepackage{amsmath}	
\usepackage{subfig}
\usepackage{caption}
\usepackage{xspace}
\usepackage{tikz}
\usetikzlibrary{shapes.geometric, arrows}

\newcommand{\unit}{{\sc unit}\xspace}
\newcommand{\orsim}{{\sc outerrim}\xspace}

\title[Non-Poissonian extensions to HOD]{Improving and extending non-Poissonian distributions for satellite galaxies sampling in HOD: applications to eBOSS ELGs}

\author[Bernhard Vos-Ginés et al.]{
Bernhard Vos-Ginés,$^{1,2,3}$\thanks{E-mail: mnras.org.uk (KTS)}
Santiago Avila$^{1,2,4}$
Violeta Gonzalez-Perez,$^{1,3}$
and Gustavo Yepes$^{1,3}$
\\
$^{1}$ Departamento de F\'isica Te\'orica,  Universidad Aut\'onoma de Madrid, 28049 Madrid, Spain \\
$^{2}$ Instituto de F\'isica Teorica UAM-CSIC, Universidad Aut\'onoma de Madrid, 28049 Cantoblanco, Madrid, Spain\\
$^{3}$ CIAFF, Facultad de Ciencias, Universidad Aut\'onoma de Madrid, 28049 Madrid, Spain \\
$^{4}$ Institut de física d'altes energies (IFAE) The Barcelona Institute of Science and Technology campus UAB, 08193 Bellaterra Barcelona, Spain
}

\date{Accepted XXX. Received YYY; in original form ZZZ}

\pubyear{2023}

\begin{document}
\label{firstpage}
\pagerange{\pageref{firstpage}--\pageref{lastpage}}
    \maketitle

\begin{abstract}
Halo Occupation Distribution (HOD) models help us to connect observations and theory, by assigning galaxies to dark matter haloes. In this work we study one of the components of HOD models: the probability distribution function (PDF), which is used to assign a discrete number of galaxies to a halo, given a mean number of galaxies. For satellite galaxies, the most commonly used PDF is a Poisson Distribution. PDFs with super-Poisson variances have also been studied, allowing for continuous values of variances. This has not been the case for sub-Poisson variances, for which only the Nearest Integer distribution, with a single variance, has been used in the past. In this work we propose a distribution based on the binomial one, which provides continuous sub-Poisson variances. We have generated mock galaxy catalogues from two dark-matter only simulations, \unit and \orsim, with HOD models assuming different PDFs. We show that the variance of the PDF for satellite galaxies affects the one-halo term of the projected correlation function, and the Count-In-Cells (CIC) one point statistics.
We fit the clustering of eBOSS Emission Line Galaxies, finding a preference for a sub-poissonian PDF, when we only vary the parameter controlling the PDF variance and the fraction of satellites. Using a mock catalogue as a reference, we have also included both the clustering and CIC to constrain the parameters of the HOD model. CIC can provide strong constraints to the PDF variance of satellite galaxies. 

\end{abstract}




\section{Introduction}
\label{sec:introduction}

The nature of dark matter and dark energy are two of the greatest mysteries in cosmology. The large scale structure of the Universe is a powerful tool to investigate these two components, ant it has gained significant attention in the last 20 years  \citep{2dfsurvey,Alam_2021,Abbott_2022}.
During this period of time, large scale surveys have increased significantly the volume of the Universe that has been mapped. Cosmological simulations must adapt to this growing volumes \citep{Heitmann_2016,Heitmann_2019,Chuang_2019,uchuu,abacus}.

In general, the largest cosmological simulations only include dark matter particles information and gravity-only evolution \citep[e.g.][]{Heitmann_2019,abacus},  whereas other smaller simulations may include gas matter particles with hydrodynamical evolution  (gas cooling, AGN feedback, etc.) as well \citep[e.g.][]{pillepich2018,Flamingo}. Thus, for the large dark-matter only simulations a connection between dark matter and tracers such as galaxies needs to be applied~\citep{Somerville,Wechsler2018}. Several methods have been developed for this purpose, such as Semi-Analytical Models, which encapsulate the processes that we understand to be most important for the formation and evolution of galaxies into coupled differential equations~\citep[e.g. ][]{Baugh2006}. A simpler model consists on Sub-halo Abundance Matching (SHAM), 
which links observed galaxies with simulated haloes, relying in a monotonic correspondence between halo mass functions and luminosity functions, with a certain level of scatter \citep[e.g.][]{favole2016,yu2023}.

An other relatively simple approach is the Halo Occupation Distribution (HOD) modelling~\citep[e.g.][]{benson2000,seljak2000,COORAY_2002,Berlind2003,zheng2005,hearin2016}. This can be applied to large simulations without the need to be complete in the subhalo mass function and without requiring resolving too much internal properties of the haloes. The parameter range can be explored fast, and they are typically used for very large cosmological simulations  \citep{Zehavi_2005,manera2013, carretero2015,Avila_2020, Yuan_2023,Rocher_2023}.

One key ingredient of HOD models is the mean HOD, which controls the expected number of galaxies to be placed in a halo of a given mass. Another key ingredient is the probability distribution function (PDF) used to sample that mean HOD. The Poisson distribution is the most commonly employed approach in the literature to place satellite galaxies within haloes \citep[e.g.][]{Zehavi_2005,carretero2015,Avila_2018,Rocher_2023}. However, PDF with sub-Poisson and super-Poisson variances have been used in the literature \citep{Jimenez_2019,Avila_2020}. In the case of super-Poisson variances, the negative binomial distribution has been used, which can parametrize the variance continuously. In the case of sub-Poisson variances, the Nearest Integer function gives a single variance, which is the smallest possible one. Nonetheless, there remains a range of variances between the Nearest Integer and Poisson distributions that has yet to be explored.

In this work we propose and validate a full prescription for the HOD PDF covering all the range in variances. The minimum variance is given by the Nearest Integer, then we cover the sub-Poissonian range with binomial distribution and an extension that we propose in \autoref{sec: PDFextensions}, we also include trivially the Poisson distribution and we cover the super-Poissonian space with an improved prescription of the Negative Binomial distribution.
Ongoing galaxy surveys such us Euclid \citep{laureijs2011euclid} or DESI \citep{desicollaboration2016} 
will be able to constrain the satellite PDF for different tracers. On the one hand, these constraints can give a unique insight to the physics of galaxy formation that rule the galaxy-halo relation. On the other hand, we will show that the PDF variance can affect severely several clustering statistics \citep[see also][]{Avila_2020}. Hence, if this is not accounted for, it may bias our cosmological interpretation when studying galaxy clustering \citep[see appendix B of][]{Avila_2020}.

In general, two-point correlation measurements are usually used for constraining HOD parameters, but in the scientific community an interest started for searching alternative statistics to constrain the HOD such as 2k-Nearest Neighbours \citet{Yuan_2023}, which have demonstrated to be at least as good as two-point statistics, providing complementary information and reducing possible degeneracies between parameters. 

In this work, we introduce Count-in-Cells (CIC) as an alternative statistics for constraining HOD and we will show that it has a very promising constraining power.

The dark matter simulations and observational data from eBOSS used in this work are described in \autoref{sec: data}. \autoref{sec: HODmodel} describes the components of the HOD model. In \autoref{sec: PDFextensions}, we introduce the binomial and extended binomial distributions and  we evaluate their performance in \autoref{sec:performance}. In \autoref{sec:clustering} we investigate the impact the probability distribution function (PDF) has on two-point statistics and we also fit the galaxy catalogues to eBOSS data in \autoref{sec: fitting_data}. In \autoref{sec:CIC} we evaluate the constraining possibilities of Counts-in-Cells and we use it to reproduce a model catalogue of galaxies in \autoref{sec:CIC_and_clustering}. Finally we summarize the results in \autoref{sec:conclusions}.

\section{Simulations and observations}
\label{sec: data}

In this work we make use of two dark matter-only simulations, \unit and \orsim  (\S~\ref{sec: Dark Matter Simulations}). The cosmology and the properties of both simulations are summarized in \autoref{tb: cosmologies}. The clustering of model galaxies obtained from these simulations is later compared with SDSS-IV/eBOSS data (\S~\ref{sec: Observational_data}). 

\subsection{Dark Matter Simulations}
\label{sec: Dark Matter Simulations}

The \unit simulations (or UNITsim, \citealt{Chuang_2019}) are full N-body dark matter-only simulations. Their initial conditions are set using the Zel\textquotesingle dovich Approximation \citep[][]{Zeldovich} provided by the \textsc{FastPM} code \citep{fastpm}. The UNITsims were implemented with the \textit{fixed \& paired} technique, where the Fourier-mode amplitudes are fixed to the ensemble-averaged power spectrum (\textit{fixed}) and two pairs of simulations were run with a $\pi$ offset on the phases within each pair (\textit{paired}) \citep{Angulo2016}. Both techniques reduce the wavemodes variance considerably, raising the effective volume of the simulations up to $V_{\rm eff} = 150$ $ \emph{h}^{-1} \rm{Mpc}$. Nevertheless, we only use one of the four \unit simulations in this work. The evolution of particles up to redshift $z = 0$ is done by \textsc{L-Gadget}, a version of the gravity solver code \textsc{Gadget2} \citep[][]{springel}. The dark matter haloes of \unit are obtained using the \textsc{rockstar} halo finder \citep{behroozi}, and halo masses are defined using the virial theorem.

The other simulation we use in this work is \orsim \citep{Heitmann_2019}. It was run using the Hardware/Hybrid Accelerated Cosmology Code (HACC) \citep{HABIB} . Haloes were obtained using the Friends-of-Friends (FoF) halo finder \citep{Davis} with a linking length of b = 0.168. FOF linking lengths define halo masses. 

\begin{table}
\centering
\begin{tabular}{r l c}
& \unit & \orsim\\ \hline
$\Omega_{\rm M} = 1-\Omega_\Lambda$ & 0.3089 & 0.2648 \\
$h = H_0 / 100 \hspace{0.8mm} \rm{ km} \hspace{0.5mm}\rm{s}^{-1} \rm{Mpc}^{-1}$ & 0.6774 & 0.7352 \\
$\sigma_8$ & 0.8147 & 0.8 \\
$n_s$ & 0.9667 & 0.963 \\
$V_{\rm box} = L_{\rm box}^3$ $\left[\rm{Gpc^3} \emph{h}^{-3}\right]$ & 1 & 27 \\
$N_{\rm P}$ & $4096^3$ & $10240^3$ \\
$m_{\rm P}$ $\left[ h^{-1}\,{\rm M}_{\sun}\right]$ & $1.25 \cdot 10^9$ & $1.85 \cdot 10^9$ 
\end{tabular} 
\caption{\label{tb: cosmologies}. \unit and \orsim simulation parameters. $\Omega_{\rm M}$ is the dimensionless dark matter density, obtained dividing the dark matter density by the critical density $\rho_{\rm crit}$. Taking into account a negligible radiation density $\Omega_r$ and the assumption of an Euclidean Universe $\Omega_k = 0$, $\Omega_{\rm M}$ is directly related to the dark energy density $\Omega_\Lambda$ by $\Omega_{\rm M} = 1-\Omega_\Lambda$. $\sigma_8$ is the amplitude of the density fluctuations; $n_s$ is the spectral index; $V_{\rm box}$ is the volume of both cubic simulations; $N_{\rm P}$ is the number of dark matter particles; and $m_{\rm P}$ the mass of dark matter particles.}
\label{tab: sim_properties}
\end{table}

Here we analyse single simulation snapshots. For \unit we consider $z = 0.8594$, which is the closest snapshot to the eBOSS ELG effective redshift $z_{\rm eff} = 0.845$ \citep{Raichoor_2020}. In the case of \orsim, the closest snapshot corresponds to $z = 0.865$. For both simulations, we only consider haloes with at least 21 particles. 

\subsubsection{The bias function}

\begin{figure*}
    \centering{}
    \subfloat[]{
        \includegraphics[width=0.48\textwidth]{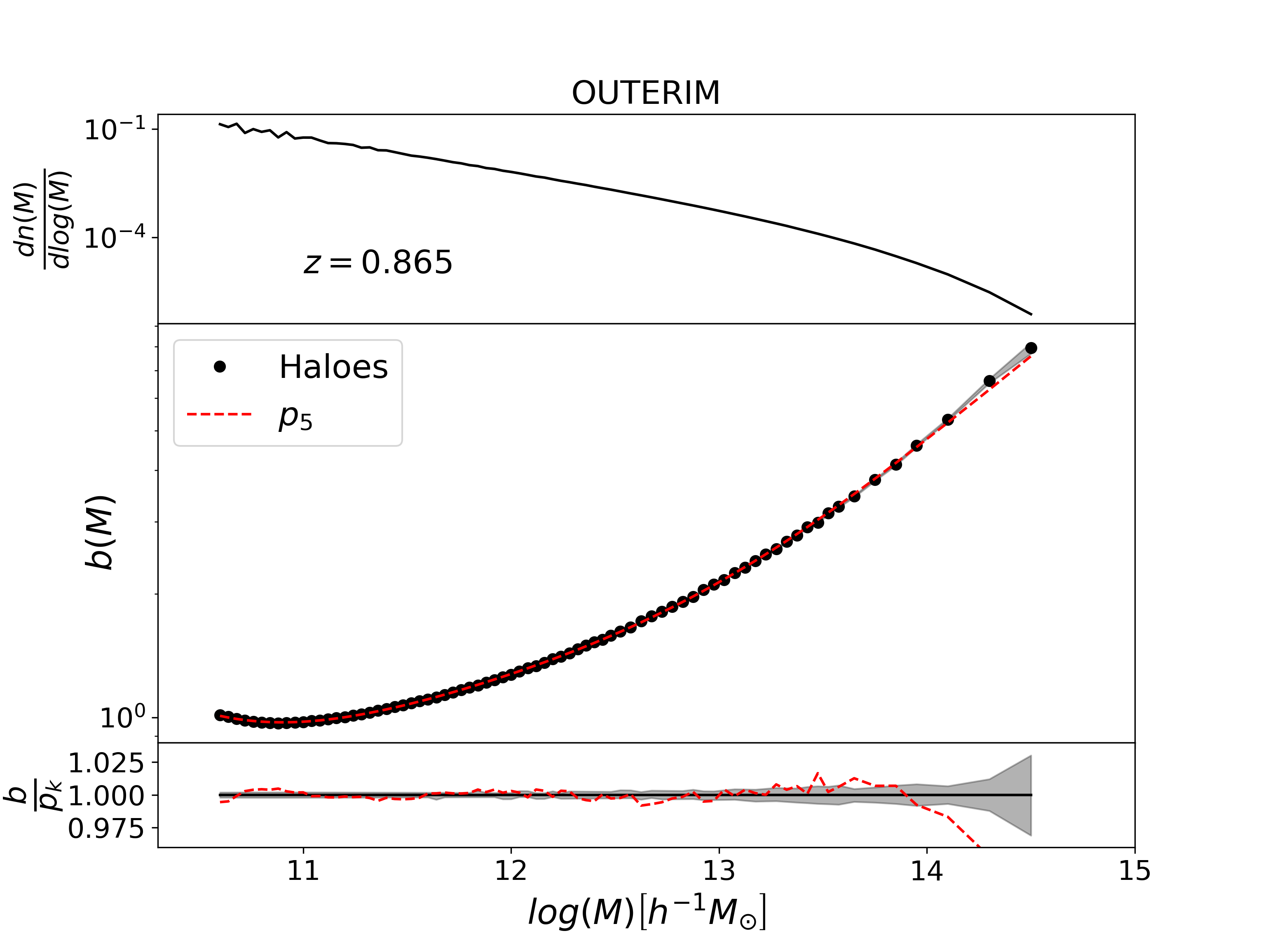}
        \label{fig:UNITSIM_BIAS_FUNCTION}}
    \subfloat[]{
        \label{fig:OUTERIM_BIAS_FUNCTION}
        \includegraphics[width=0.48\textwidth]{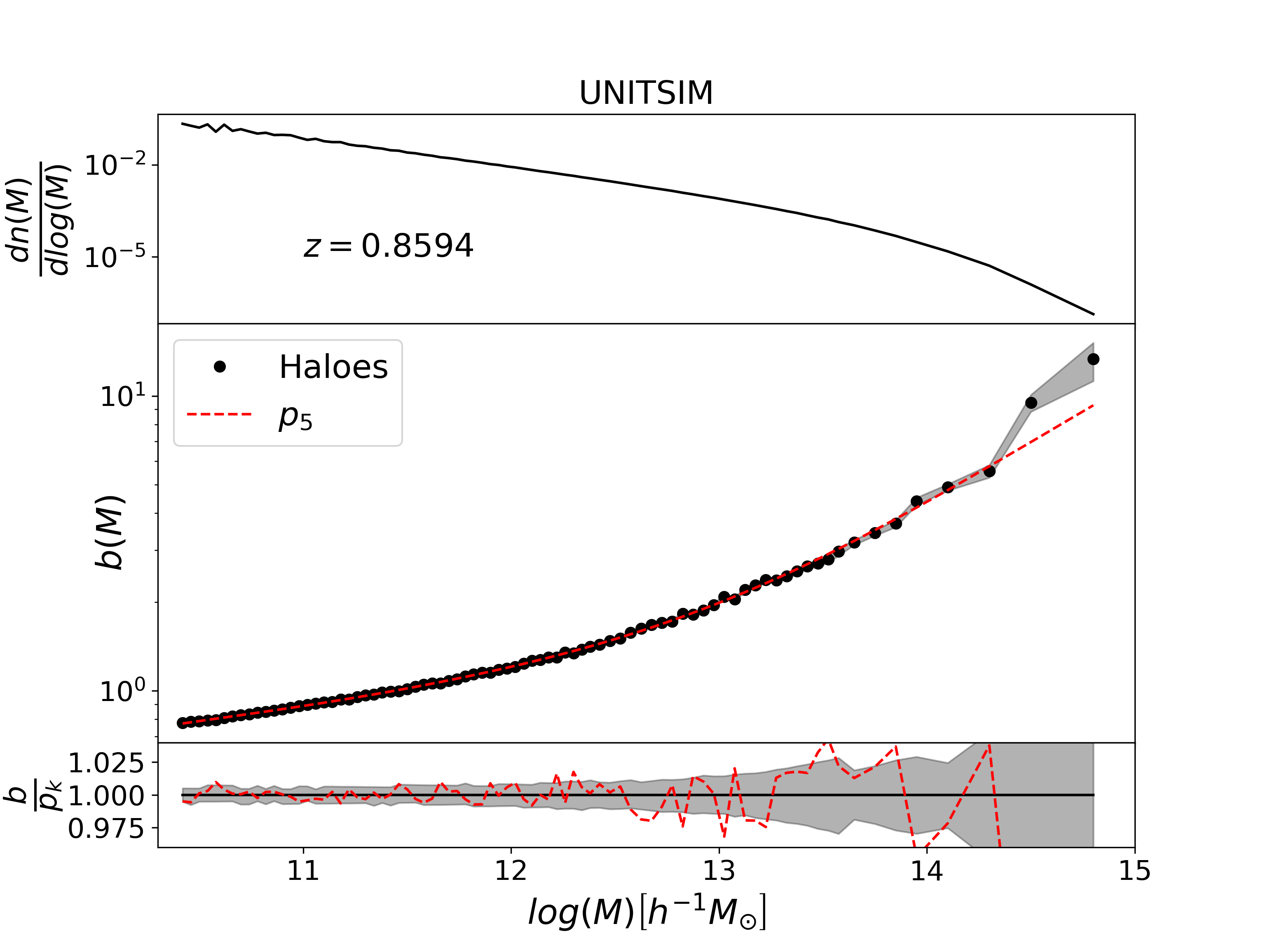}}      
\caption{{\it Left}: at the top we represent the \unit halo mass function, in the center the halo bias function (black points) with $1 \sigma$ confidence interval calculated from $\Delta \chi(b)^2 = 1$ and a fifth order polynomial (red dashed line) fitting the halo bias function. In the bottom we show the ratio between the polynomial and the halo bias function. {\it Right}: same for \orsim simulation. We find around 1 per cent differences in the halo bias function, when comparing \orsim case with Figure 1 of \citet{Avila_2020}.} 
\label{fig:BIAS_FUNCTION}
\end{figure*}

Following \citet{Avila_2020}, we first compute the bias and halo mass functions, which are used to set constrains in the HOD model as we will see in \autoref{sec:mean_HOD}. For both functions we use halo mass bins that are narrow for small halo masses and become wider for higher masses to account for the lower number of haloes (see \autoref{fig:BIAS_FUNCTION}). In \autoref{sec:mean_HOD} we relate those quantities to the HOD parameters. 

We calculate the halo bias function for \orsim following the same procedure as in \citet{Avila_2020}, in order to allow a direct comparison. We find differences in the halo bias function below 1 per cent with respect to \citet{Avila_2020}.

However, in the case of \unit, we obtain the halo bias function slightly different, by computing the power spectrum for each mass bin. We find this method provides more stable results than when considering the two-point correlation function used in \citet{Avila_2020}. For obtaining the bias from the power spectrum, we use the following limits $k_{\rm min} = 2\pi /L_{\rm box} + dk/2$ and $k_{\rm max} = 2\pi N_{\rm grid}/L_{\rm box} + dk/2$, with linear spacing $dk = 2\pi /L_{\rm box}$ and $N_{\rm grid} = 512$. For each mass bin we obtain the correspondent bias minimizing the $\chi^2$ function:

\begin{equation}
\label{eqn:1}
    \chi_i ^2(b) = \sum_{k=k_{\rm min}}^{k_{\rm cut}} \frac{\left(b^2 P_{\rm th}(k)- P_{h,i}(k)\right)^2}{\Delta P_{h,i}(k)^2} \hspace{2mm},
\end{equation}
where $P_{\rm th}(k)$ is the linear matter power spectrum computed from \unit dark matter particles, $P_{h,i}(k)$ is the halo power spectra for each mass bin $i$ and $b$ is the linear bias. We use $k_{\rm cut} = 0.1$ $\rm{Mpc^{-1}} \emph{h}$, since the effect of non-linearities is non-negligible beyond this scale. The error in the power spectrum $\Delta P_{h,i}(k)^2$ has the following expression:

\begin{equation}
    \Delta P_{h,i}(k)^2 = \frac{\left(2\pi\right)^2}{k^2 dk V_{\rm box}} \left(P_{h,i}(k)+\frac{1}{n_i}\right)^2 \hspace{2mm},
\end{equation}
with the halo number density is $n_i = N_{h,i}/V_{\rm box}$ $\left[\emph{h}^{-1} \rm{Mpc}\right]^{-3}$ at the mass bin $i$, and $N_{h,i}$ as the number of haloes at that mass bin. 

We finally obtain the linear bias for each halo mass bin considering $\chi_{\rm min} ^2$ with a $1 \sigma$ confidence interval given by $\Delta \chi ^2 = 1$.  In \autoref{fig:BIAS_FUNCTION} we represent the halo mass and the halo bias functions for \orsim and \unit simulations. Higher-mass haloes are less frequent and they also have a higher bias. 

For both simulations we fit the resulting bias function with a fifth order polynomial (represented by a red dashed line in \autoref{fig:BIAS_FUNCTION}). The polynomial fit is able to encapsulate the simulation measurements within $1 \sigma$, except for the two most massive bins of the \unit simulation. Nevertheless, we expect this effect to be negligible, as very few haloes are found within those two massive bins.

\subsection{Observational data}
\label{sec: Observational_data}

In this work we generate mock catalogues with the number density and linear bias fixed to the Emission-Line Galaxies (ELG) from the extended Baryon Oscillation Spectroscopic Survey (eBOSS)  from the DR16 Sloan Sky Digital Survey (SDSS) IV \citep{Dawson_2016, Ross2020, Alam_2021}. These galaxies have redshifts in the range $0.6 < z < 1.1$. eBOSS ELGs are mostly star-forming galaxies with strong spectral emission lines that enable a fast and reliable determination of their redshift \citep{Raichoor_2020}.

We use HOD models to populate with galaxies the \unit and \orsim simulations, which assume different cosmologies (see \autoref{tab: sim_properties}). Below we explain how we obtain the needed number density and linear bias from the observations.

\subsubsection{Number density}

The number density, $n_{\rm gal}$, is the total number of observed galaxies per observed volume. To take into account the incompleteness due to the photometric target selection, redshift failures and other effects such as fiber collision, a weight is associated with each galaxy \citep{Raichoor_2020,Ross2020}. In order to compute the total number of observed galaxies we take into account those weights.

Since the comoving survey volume depends on the assumed cosmology, $n_{\rm gal}$ varies with cosmology. The survey volume is calculated as follows:

\begin{equation}
    V_{\rm eBOSS} = \frac{1}{3} \left(\chi(z_{\rm max})^3-\chi(z_{\rm min})^3\right)\cdot A_{\rm eff} \left(\frac{\pi}{180 \rm deg}\right)^2,
\end{equation}

where $\chi(z)$ is the comoving distance, which also depends on cosmology. $A_{\rm eff} = A_N + A_S$, with $A_{N(S)} = 369.451(357.546)$ $\rm{deg}^2$ is defined as the effective areas for the North (South) Galactic caps covered by eBOSS. Finally, $z_{\rm min}=0.6$  and $z_{\rm max}=1.1$ are the minimum and maximum redshifts of the observed ELGs considered in the LSS catalogues  \citep{Raichoor_2020}. 

We compute the volumes of the different galactic caps and the total volume of the eBOSS ELGs: $V_{\rm N}= 0.425 (0.467)$ $\left[ \emph{h}^{-1} \rm{Gpc}\right]^3$, $V_{\rm S} = 0.411(0.452)$ $\left[ \emph{h}^{-1} \rm{Gpc}\right]^3$ and $V_{\rm eBOSS} = 0.836(0.919)$ $\left[ \emph{h}^{-1} \rm{Gpc}\right]^3$ considering \unit (\orsim) cosmologies.

We obtain $n_{\rm gal} = 2.406 \cdot 10^{-4} $ $\left[\rm{Mpc^{-1} \emph{h}}\right]^3$ for \unit and $n_{\rm gal} = 2.187 \cdot 10^{-4}$ $\left[\rm{Mpc^{-1} \emph{h}}\right]^3$ for \orsim cosmologies (\autoref{tab: sim_properties}).

\subsubsection{Linear Bias}

The linear bias, $b_{\rm gal}$, can be defined as the ratio between the overdensity of galaxies and the overdensity of dark matter at large scales. We calculate the bias using the Kaiser factor \citep{kaiser} to relate the observed galaxies and the dark matter monopoles $\xi_0$ while accounting for Redshift Space Distortions (RSD). 

\begin{equation}
    \xi_0 (s) = \left(b_{\rm gal}^2+\frac{2}{3} b_{\rm gal}f + \frac{1}{5}f^2\right) \cdot \xi_{\rm th} (s) ,
\end{equation}
we use the approximation $f(z) = \left(\Omega_{\rm M} (z)\right)^{0.545} $ \citep{peebles1980,Linder_2005} for the growth rate of structure, evaluated at $z_{\rm eff} = 0.845$, and $\xi_{\rm th}$ is the monopole of the matter two-point correlation function. For both simulations we consider the range: $15 < s < 75$ $\emph{h}^{-1}\rm{Mpc}$, with linear binning: $\Delta s = 5$ $\emph{h}^{-1}\rm{Mpc}$. For the \unit simulation we calculate the monopole of dark matter particles for 0.5 per cent of the total particles \footnote{The differences between using $0.05$ and $0.5 $ per cent of the total number of particles is below $3 $ per cent in this range }. For the \orsim simulation, we use the monopole provided by \textsc{camb} linear theory evaluated at the \orsim cosmology since we do not have enough information on the particles.
The monopoles from \unit \textsc{camb} linear theory and dark matter particles are compatible for scales $s = 15-75$ $\emph{h}^{-1}\rm{Mpc}$, with errors below 4 per cent.

Finally, for the \unit simulation we need to include the simulation and survey volume ratio to correct the amplitude of the precision matrix in the $\chi^2$ function.  We compute the $\chi^2$ as follows: 

\begin{equation}
    \chi ^2(b) = \sum_{s,s'} \left(\xi_{0} (s)-\xi_{0,d} (s)\right)^T C^{-1} (s,s')  \left(\xi_{0}(s')-\xi_{0,d} (s')\right) \hspace{2mm}  ,
\end{equation}
with $V_{\rm box}$ the volume of the simulation and $C^{-1}(s,s')$ is the inverse of the covariance or precision matrix. The covariance matrix is calculated using $N_{\rm EZ} = 1000$ Effective Zeldovich (EZ) Mocks \citep{Zhao_2021} for both simulations. The bias is found by minimising the $\chi^2$, with a $1 \sigma$ confidence interval computed using $\Delta \chi^2 = 1$.  

We obtain $b_{\rm gal} = 1.37 \pm 0.03 $ for \unit and $b_{\rm gal} = 1.33 \pm 0.02$ for \orsim cosmologies.

\section{Model galaxies} 
\label{sec: HODmodel}

Halo Occupation Distribution (HOD) models populate dark matter haloes at a given redshift with galaxies using analytical equations that relate the probability of finding a galaxy of a certain type with the mass of the host halo. We distinguish between two types of galaxies: centrals and satellites. 

Central galaxies are placed at the center of host haloes and share their velocity. Due to this definition, a particular halo will not host more than one central galaxy. 

On the other hand, satellite galaxies do not have to share the position and velocity of its host halo. They are associated to haloes with a given radial and velocity profile. 

For a generic HOD model, equations are chosen to describe the following properties:

\begin{itemize}
    \item Shape of the mean Halo Occupation Distribution (HOD) (\S~\ref{sec:mean_HOD}).
    \item Probability distribution function (PDF) for both the number of central and satellite galaxies within a halo (\S~\ref{sec:PDF}). We will extend the satellite PDF in \autoref{sec: PDFextensions}.
    \item Spatial and velocity distribution of satellites within haloes (\S~\ref{sec:spatial_and_velocity}).
\end{itemize}

\subsection{Mean Halo Occupation Distribution}
\label{sec:mean_HOD}

The shape of the mean HOD, parametrises how many galaxies will be hosted, on average, by a halo of a given mass. Usually, two analytical expressions for the shape of the mean HOD distribution are used: one for satellites and another for centrals. 

Those expressions for the mean distribution of galaxies can be used to calculate the total number density and bias of the galaxy catalogue resulting from an HOD, given we know the halo mass function and halo bias functions (see \autoref{sec: Dark Matter Simulations}):

\begin{equation}
\label{eqn: number_density_galaxies}
    n_{\rm gal} = \int \frac{dn(M)}{dlog M} \left[\langle N_{\rm cen} (M)\rangle+\langle N_{\rm sat} (M) \rangle \right]  d \rm{log}M
\end{equation}

\begin{equation}
\label{eqn: bias_galaxies}
    b_{\rm gal} =\frac{1}{n_{\rm gal}} \int \frac{dn(M)}{dlog M} b(M) \left[ \langle N_{\rm cen} (M) \rangle+\langle N_{\rm sat} (M)\rangle\right] d \rm{log} M
\end{equation}

We can also calculate the fraction of satellites as follows:

\begin{equation}
\label{eqn: fraction_of_satellites_galaxies}
    f_{\rm sat} =\frac{1}{n_{\rm gal}} \int \frac{dn(M)}{dlog M} \langle N_{\rm sat} (M) \rangle d \rm{log} M \hspace{2mm}
    {.}
\end{equation}

We fix the number density of galaxies and the galaxy bias to the observed values of eBOSS ELG data calculated in \autoref{sec: Observational_data}. The fraction of satellites is set as free parameter.

In this work we follow the assumptions presented in \citep{Avila_2020} for the modelling of satellite and central galaxies.

Central galaxies follow an asymmetric Gaussian distribution. This description is motivated by semi-analytical models of galaxy formation and evolution \citep[e.g.][]{Gonzalez_Perez_2017}. For central galaxies we assume the following shape: 

\begin{equation}
\label{eqn: Ncen_mean}
    \langle N_{\rm cen} (M) \rangle = \frac{A_c}{\sqrt{2\pi}\sigma} 
    \times \left\lbrace\begin{array}{c}  e^{-\frac{\left(\log{M}-\mu\right)^2}{2\sigma^2}} \hspace{3mm} \log{M} \leq \mu\\ \left(\frac{M}{10^\mu}\right)^{\gamma} \hspace{8mm} \log{M} \geq \mu\end{array}\right. \hspace{2mm}
    {,}
\end{equation}
where $M$ is the halo mass and $A_c$ determines the amplitude of the Gaussian, which has mean $\mu$ and variance $\sigma^2$. For $\log{M} \geq \mu$, $\gamma < 0$ controls the sharpness of the decaying power law. 

For satellite galaxies, we assume they follow a power-law: 

\begin{equation}
\label{eqn: Nsat_mean}
    \langle N_{\rm sat} (M) \rangle = A_s \cdot \left(\frac{M-M_0}{M_1}\right)^\alpha \hspace{4mm} M >M_0 \hspace{2mm},
\end{equation}
where $A_s$ controls the fraction of satellites and $\alpha > 0$ provides how steep is the power law. The average number of satellites is zero if $M \leq M_0$ and it increases as the halo mass increases starting from this point. If $A_s = 1$ and $M_0 \ll M_1$, $M_1$ represent the mass of haloes in which we expect 1 galaxy satellite on average. In this work we fix the mass parameters $M_0$ and $M_1$ with a relation to $\mu$ given by $\rm{log} \left(M_0\right) = \mu - 0.05$ and $\rm{log} \left(M_1\right) = \mu + 0.35$ \citep{Gonzalez_Perez_2017}. We can see the shape of the average number of satellites per halo mass in the black solid line of \autoref{fig:HOD_scatter}.

HOD parameters $\alpha$, $\sigma$ and $\gamma$ are fixed throughout this work to the values shown in \autoref{tab: default_HOD}. $\mu$,$A_c$,$A_s$ are determined by the number density $n_{\rm gal}$, the linear bias $b_{\rm gal}$ from eBOSS ELG observations and also the fraction of satellites $f_{\rm sat}$ (free parameter of our model), using \autoref{eqn: number_density_galaxies}, \autoref{eqn: bias_galaxies} and \autoref{eqn: fraction_of_satellites_galaxies}, and the latter. We define a default HOD in \autoref{tab: default_HOD} that will be used several times in this work, in which we fix $n_{\rm gal} = 6 \cdot n_{\rm eBOSS}$, $b_{\rm gal} = 1.37$ and $f_{\rm sat} = 0.3$.

\subsection{Probability distribution function}
\label{sec:PDF}

In order to place galaxies into haloes we need to use a discrete probability distribution function (PDF) that will determine how many galaxies of a given type, $N$, we place given a mean number $\langle N\rangle$ determined by the mean HOD described above. 

For all discrete probability distribution functions considered in this work, the number of central or satellite galaxies each halo will host is determined by drawing a random number $\theta \in [0,1)$. Then, using the cumulative probability distribution function $P_{\rm C}(N)= \sum^{N}_{x} P(x)$, we determine the value $\zeta$ such that $P_{\rm C}(N = \zeta)<\theta$ and $P_{\rm C}(N = \zeta+1)>\theta$. Then, the number of central or satellite galaxies hosted by the halo are $N = \zeta$.

In the case of galaxy centrals, the Nearest Integer Distribution is always used as haloes can host either one or none central galaxy (for mean values between 0 and 1, the Nearest Integer distribution is identical to a Bernouilli distribution). 

In the case of satellite galaxies, PDFs with different variances have been used in the literature: Poisson distribution, the standard case for most HOD models (\S\ref{sec:poisson}), super-Poisson through the Negative Binomial distribution (NB, \S\ref{sec:super-poisson}) and sub-Poisson. So far in the literature, for this last case a Nearest Integer (NI) Distribution (\S\ref{sec:sub-poisson}) has been used. However, this involves a single value for the sub-Poissonian variance. We will introduce in \autoref{sec: PDFextensions} two more PDFs that increase the flexibility when assigning satellites to dark matter haloes: the Binomial distribution (\S\ref{sec: binomial}) and an extended Binomial distribution (\S\ref{sec:beyond_binomial}). This is central focus of this work.

Some examples of non-Poisson PDFs in the literature are described here.
\citet{Jimenez_2019} uses a super-Poissonian PDF for satellites considering a Negative binomial distribution (NB). This was motivated for model star forming galaxies. Although a Nearest Integer distribution (NI) is used only for centrals, \citet{Berlind2003} found NI in better agreement with galaxies selected by Smoothed Particle Hydrodynamics (SPH) and Semi-Analytic models, independently of them being centrals or satellites. Several best fits on eBOSS data found in \citet{Avila_2020} also show preference for NI and NB. However, there is a significant gap in the variance of the PDF from the Poisson distribution to the only sub-Poisson function considered (NI), motivating the modelling of a continuous distribution.

In this work we quantify the deviations of the PDF variance with respect to that of a Poisson distribution by the parameter $\omega_{\sigma}$:

\begin{equation}
\label{eqn:variance_omega}
    \sigma^2 \equiv \lambda \left(1+\omega_{\sigma} \lambda\right)
\end{equation}

The parameter $\omega_{\sigma}$ is defined differently for super and sub-Poisson variances, as it is detailed below and in \autoref{sec: PDFextensions}. \autoref{eqn:variance_omega} is adequate for a range of variances that can be continuous. Nevertheless, as it is detailed in \autoref{sec:sub-poisson}, the Nearest Integer distribution is the only distribution considered in this work that has a single variance and its expression differs from the above equation.

\subsubsection{Poisson Distribution}\label{sec:poisson}

The most commonly used probability distribution to determine how many satellite galaxies are placed in a particular halo of a given mass is the Poisson distribution. A Poisson PDF can be written as follows for a halo with an average satellite galaxy $\lambda = \langle N_{\rm sat}  \rangle$: 

\begin{equation}
\label{eqn:Poisson}
    P (N_{\rm sat} ; \lambda )= \frac{e^{-N_{\rm sat}} \lambda^{N_{\rm sat}}}{N_{\rm sat}!}
\end{equation}
For this distribution, the variance is equal to the mean  $\sigma^2 = \lambda = \langle N_{\rm sat}  \rangle$. The shaded black region in \autoref{fig:HOD_scatter} represents the theoretical Poisson variance for a particular HOD model, and the green lines represent the observed variance after 20 realizations of galaxy catalogues, which are computed with a Poisson distribution. As expected, the black shaded region and green lines are in agreement.

\subsubsection{Super-Poisson distribution: negative binomial}
\label{sec:super-poisson}

The probability of getting an integer random variable, $N_{\rm sat}$, for the negative binomial (NB) PDF can be written as follows:

\begin{equation}
\label{eqn:NegBinomial}
    \rm{NB} (N_{\rm sat}; p,q) = \frac{\Gamma(N_{\rm sat}+q)}{\Gamma(q)\Gamma(N_{\rm sat}+1)} p^q \left(1-p\right)^{N_{\rm sat}} ,
\end{equation}
where $q$ traditionally describes the number of successes, hence is defined as a natural number. This quantity, $q$, can be naturally extended to positive real numbers despite losing its original meaning: $q \in \mathbb{R}^+$. $0 < p < 1$ is the probability of success and $N_{\rm sat}$ the number of failures, which is the random variable of this PDF. In that traditional interpretation $N_{\rm sat}+q-1$ would represent the total number of trials, from which $q$ can inherit the nature of a free parameter. In the context of HOD models, $N_{\rm sat}$ is the number of satellite galaxies, and $p$ and $q$ are the PDF parameters that determine its mean,
\begin{equation}
\label{eqn: mean_neg_binomial}   
     \lambda \equiv \langle N_{\rm sat} \rangle = \frac{pq}{1-p} \,,
\end{equation}
and standard deviation, 
\begin{equation}
\label{eqn: std_neg_binomial}   
      \sigma^2 = \lambda \left(1+\frac{\lambda }{q}\right) \equiv \lambda \left(1+\omega_{\sigma} \lambda\right) \, .
\end{equation}

In this case, the parameter that controls the variance deviations with respect to Poisson, $\omega_{\rm \sigma}$, is defined as $\omega_{\rm \sigma} = \frac{1}{q} > 0$ to parameterise.

In the limit $q \xrightarrow[]{} \infty$ we recover the Poisson distribution, corresponding to $\omega_{\rm \sigma} = 0$. \autoref{eqn:variance_omega} also represent the variance of the binomial distribution (See \autoref{sec: binomial}) and the extended binomial distribution (See \autoref{sec:beyond_binomial}) for $\omega_{\rm \sigma} < 0$.

As we will see in \autoref{sec: binomial} our code will be implemented with a free parameter $\omega$, which although it is motivated to follow the behaviour of $\omega_{\sigma}$ in \autoref{eqn:variance_omega} for the negative binomial distribution, in some parts of our parameter space using sub-Poissonian distributions this is not possible. We represent the $1-\sigma $ contours using the negative binomial distribution for $\omega = 0.5$ \footnote{Note that we use $\omega$ instead of $\omega_{\rm \sigma}$, since we are referring to the input for the HOD model, which in principle follow the behaviour of $\omega_{\rm \sigma}$ except in some small region of the parameter space (see \autoref{sec:correctionbinomial})} in the blue lines of \autoref{fig:HOD_scatter}, computed using 20 galaxy catalogue realizations.

The negative binomial distribution used in \citet{Jimenez_2019} and \citet{Avila_2020} had a slightly different parametrization of the variance, with $\sigma^2 = \lambda \left(1+\beta \right)$ where $\beta = \omega_{\rm \sigma} \lambda$.

\subsubsection{Sub-Poisson distribution: Nearest Integer}\label{sec:sub-poisson}

\begin{figure}
    \centering
    \includegraphics[width=0.48\textwidth]{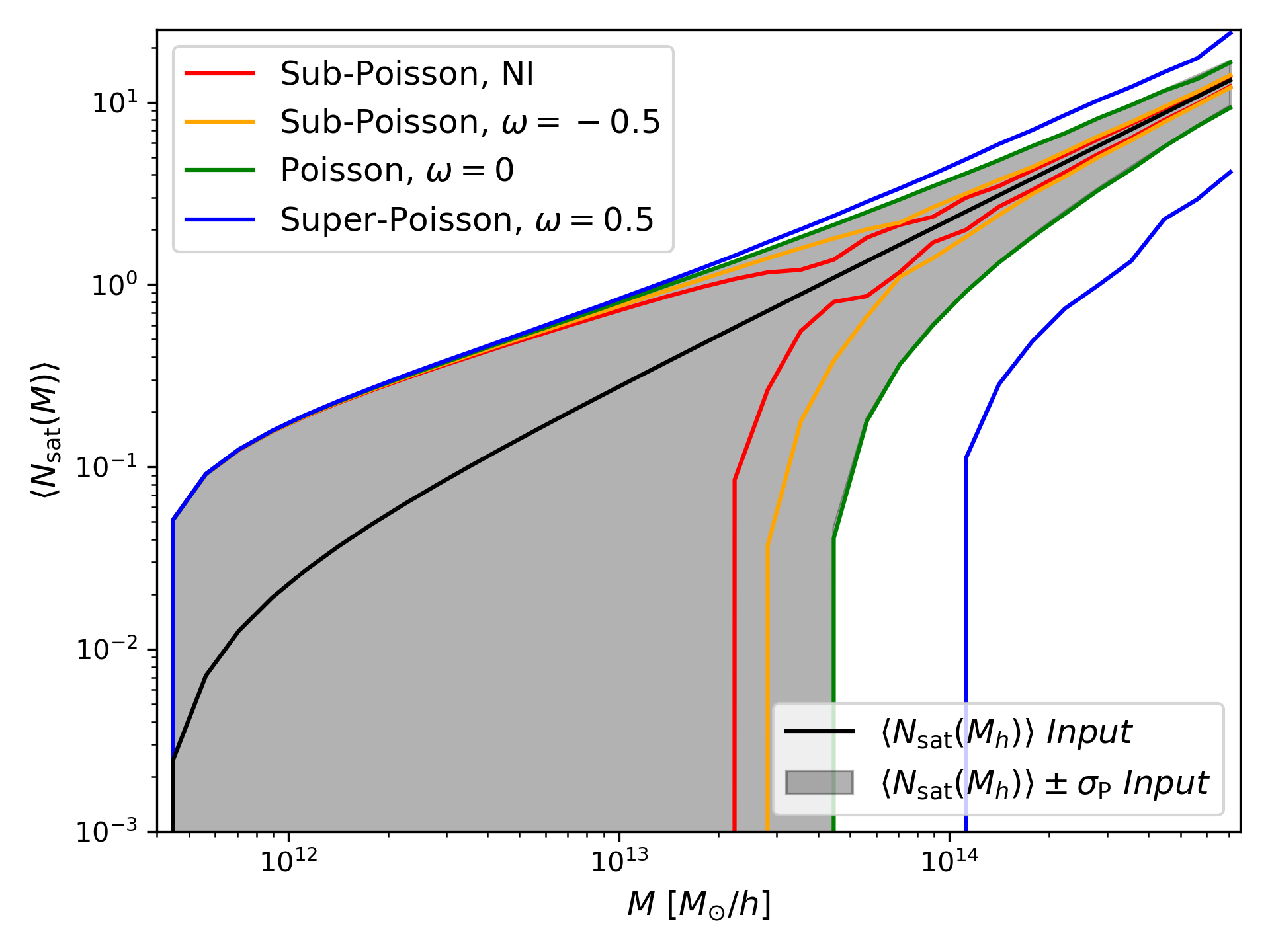}
    \caption{Mean and standard deviation of satellite galaxies as a function of halo mass $M$. The solid black line shows the mean default HOD, and the black 
    shade is the theoretical $1 \sigma$ contour considering a Poisson distribution ($\sigma = \sigma_{\rm P}$). Solid colored lines represent the measured $1 \sigma$ standard deviation around the mean, 
    for different values of $\omega$. }
    \label{fig:HOD_scatter}
\end{figure}

The probability of getting an integer random variable, $N_{\rm sat}$, with $\lambda = \langle N_{\rm sat}\rangle$, for the Nearest Integer (NI) PDF is:

\begin{equation}
\label{eqn: NI}
    \rm{NI}(N_{\rm sat};\lambda) = \left\lbrace\begin{array}{c} 1- \left(\lambda - {\rm trunc}(\lambda)\right)  \hspace{5mm} \rm{if} \hspace{5mm}  N_{\rm sat} = {\rm trunc}(\lambda)\\ \hspace{4mm} \lambda - {\rm trunc}(\lambda) \hspace{6mm} \rm{if} \hspace{6mm} N_{\rm sat} = {\rm trunc}(\lambda)+1 \\ 0 \hspace{27mm} \rm{otherwise}\end{array}\right. \hspace{2mm} ,
\end{equation}
in which $trunc(\lambda)$ is the closest lower integer to $\lambda$. The variance of this distribution is:

\begin{equation}
    \label{eqn: std_NI}
    \sigma^2 = \lambda' \left(1-\lambda'\right) \hspace{1mm},
\end{equation}
with $\lambda' = \lambda-{\rm trunc}(\lambda)$, which is the smallest possible variance. This PDF was the only sub-Poisson PDF considered in \citet{Avila_2020}, with the drawback that there is only one possible variance for each value of $\lambda$. This implies that $\sigma^2$ is not an independent parameter from $\lambda$, in contrast with the Negative Binomial case (which was already used in \citet{Avila_2020} as a super Poisson PDF.) ( \autoref{eqn: std_neg_binomial}). We represent the Nearest Integer 1$\sigma$ deviations by red solid lines in \autoref{fig:HOD_scatter}, computed using 20 galaxy catalogue realizations. The Nearest Integer Distribution is also represented by a black dashed line in \autoref{fig:std_binomial}. We note that in the case when  $0 < \langle N \rangle < 1$, the NI function is identical to the Bernouilli function, often quoted in HOD models.

\subsection{Spatial and velocity distributions of satellites}
\label{sec:spatial_and_velocity}

In this work, positions of satellite galaxies within haloes are assigned following a Navaro-Frenk and White (NFW) radial profile \citep*{Navarro_1997}. The velocity distribution of satellite galaxies are calculated considering the virial theorem, following \citet{Bryan_1998}. The implementation of these two components is described in detail in \citet{Avila_2020}.

\begin{table}
\centering
  \begin{tabular}{c |c |c |c |c |c}
 $\mu$ & $A_c$ & $A_s$ & $\alpha$  & $\sigma$ & $\gamma$ \\
 \hline
 $11.648$ & $0.0368$ & $0.03583$ & $0.9$ & $0.12$ & $-1.4$ \\

  \end{tabular}
\caption{Default mean HOD used in this work. The parameters $\mu$, $A_c$ and $A_s$ are obtained considering $f_{\rm sat} = 0.3$, $b_{\rm gal}=1.37$ and $n_{\rm gal} = 6\cdot n_{\rm eBOSS}$, in the context of \unit simulation.}
\label{tab: default_HOD}
\end{table}

\section{Extensions to the satellite PDF}
\label{sec: PDFextensions}

So far, in the literature, sub-Poisson variances for satellite galaxies have been only described with a Nearest Integer (NI) PDF \citep{Berlind2003,zheng2005,Jimenez_2019,Avila_2020}. The NI PDF only admits one possible variance which is the smallest possible. This limits the parameter space that can be explored to find the best description of galaxies. 
In this section we introduce the extensions proposed in this work to sample sub-Poisson variances continuously. We also propose a solution to mitigate numerical errors affecting non-Poisson distributions in general, derived from $\Gamma$ functions with large arguments.

\subsection{Binomial distribution}
\label{sec: binomial}

The binomial distribution, ${\rm B}$, provides a discrete range of possible sub-Poissonian 
variances. For this distribution, the probability of getting a certain number of satellite galaxies in a halo, $N_{\rm sat}$, which is the random variable of the PDF, is given by: 

\begin{equation}
\label{eqn:binomial}
     \rm{B} (N_{\rm sat}; q,p) = \frac{\Gamma \left(q+1\right)}{\Gamma \left(N_{\rm sat}+1\right)\Gamma \left(q+1-N_{\rm sat}\right)} p^{N_{\rm sat}} \left(1-p\right)^{q-N_{\rm sat}} \, ,
\end{equation}
with the parameter $q \in \mathbb{N}$ defined traditionally as the number of trials, which is the maximum number of satellites that can be placed with non-zero probability, and $0<p<1$ the probability of success, i.e. of actually placing a satellite galaxy in a halo. The possible number of satellite galaxies has to be less than the number of trials, $N_{\rm sat} \leq q$.  Unlike in the negative binomial distribution, the possible values of $q$ cannot be extended to real numbers, since negative probabilities could arise.  

Given the parameters $p$ and $q$, the mean of the binomial distribution, ${\rm B(N_{\rm sat}; q,p)}$, is:

\begin{equation}
\label{eqn: mean_binomial}
    \lambda \equiv \langle N_{\rm sat} \rangle = qp  \,.
\end{equation}

As we can see in \autoref{fig:HOD_scatter} and \autoref{eqn: Nsat_mean}, in our model, the mean number of satellite galaxies $\lambda = \langle N_{\rm sat}(M)\rangle$ increases with the halo mass, $M$. The variance of the binomial distribution is: 

\begin{equation}
\label{eqn: std_binomial}
\sigma^2 = \lambda \left(1-\frac{\lambda}{q}\right)  \equiv \lambda \left(1+\omega_{\sigma} \lambda\right)\, ;
\end{equation}

Here we define $\omega_{\rm \sigma} \equiv -\frac{1}{q}$   as the parameter to control the variances lower than Poisson \footnote{ \citet{Avila_2020} and \citet{Jimenez_2019} use another parametrization of the variance: $\sigma^2 = \lambda \left(1 + \beta\right)$ when they consider the negative binomial distribution. In the context of the binomial distribution, the possible values of $\beta$ are limited to $-\beta < \lambda (M)$, since there are no mathematically possible variances lower than the Nearest Integer variance: $\sigma^2 = \lambda \left(1-\lambda\right)$, in which $\lambda$ is an increasing function of the halo mass. Then, $\beta$ is not a proper parameter to parametrize binomial variances for the entire HOD catalogue with $\langle N(M) \rangle$, since its values have a halo-mass dependent limitation. We use instead $\omega_{\rm \sigma} = \beta / \lambda$. Now, $\omega_{\rm \sigma}$ has a constant limit from below ($-\omega_{\rm \sigma} < 1$) for all halo masses.}, following what we have previously done for the negative binomial distribution, ${\rm NB}$. Unlike the Nearest Integer distribution, the binomial distribution has sub-Poissonian variances that not only depend on the mean number of satellites, $\langle N_{\rm sat} \rangle$, but also on $\omega_\sigma$. As it happened for the negative binomial distribution, in the limit $q \xrightarrow[]{} \infty$ we recover the Poisson distribution $\left(\omega_{\rm \sigma} = 0\right)$.

The variance of the distribution has exactly the same expression as in \autoref{eqn:variance_omega}. However, since for the binomial distribution $q \in \mathbb{N}$, the only possible input values of $\omega_{\rm \sigma}$ are discrete: $\omega_{\rm \sigma} \in \left[-1,-\frac{1}{2},-\frac{1}{3},...,-\frac{1}{\infty} = 0 \right)$. We extend this range by introducing a new PDF in \autoref{sec:beyond_binomial}.

The skewness of the binomial distribution has the following expression: 
\begin{equation}
\label{eqn:skewness}
    k_3 = \frac{\left(1-\frac{2\lambda}{q}\right)}{\sigma} \, .
\end{equation}

We have provided expressions for the third first moments of the binomial distribution: the mean (\autoref{eqn: mean_binomial}), the variance (\autoref{eqn: std_binomial}), and the skewness (\autoref{eqn:skewness}). These first three moments are properly defined with the above equations for $q  \geq 1$, $q \geq 2$ and $q \geq 3$, respectively. That is, if $q < k$, the $k$-th moment of the distribution may follow another expression.

So far, we have introduced $\omega_{\rm \sigma}$ to parametrize super-Poisson (\autoref{sec:super-poisson}) and sub-Poisson (\autoref{sec: binomial}) variances. Now we need to introduce $\omega$, which will be the parameter that is input to the HOD model to control the variance of the satellite PDF. This parameter follows the behaviour of $\omega_{\rm \sigma}$, however for some regions in the parameter space $\left(\lambda > 1,\omega < -\frac{1}{\lambda}\right)$ negative probabilities arise from \autoref{eqn:binomial} for certain $N_{\rm sat}$ (see red shaded area in \autoref{fig:range_binomial}). We describe and address this limitation below.

\subsubsection{Extension to avoid unphysical negative probabilities} 
\label{sec:correctionbinomial}

Since the mean number of satellites increases with halo mass, massive enough haloes will be able to host several satellite galaxies, $\langle N_{\rm sat} \rangle > 1$. The exact number of haloes that fulfill $\langle N_{\rm sat} \rangle > 1$ also depends on other HOD parameters, such as the fraction of satellites $f_{\rm sat}$. 

If we consider our default HOD presented in \autoref{tab: default_HOD}, which corresponds to a fraction of satellites $f_{\rm sat} = 0.3$, we have that $3.4$ ($11.2$) per cent of all galaxies (satellite galaxies) are attached to haloes with $\langle N_{\rm sat} \rangle > 1$. 

\begin{figure}
    \centering
    \includegraphics[width=0.5\textwidth]{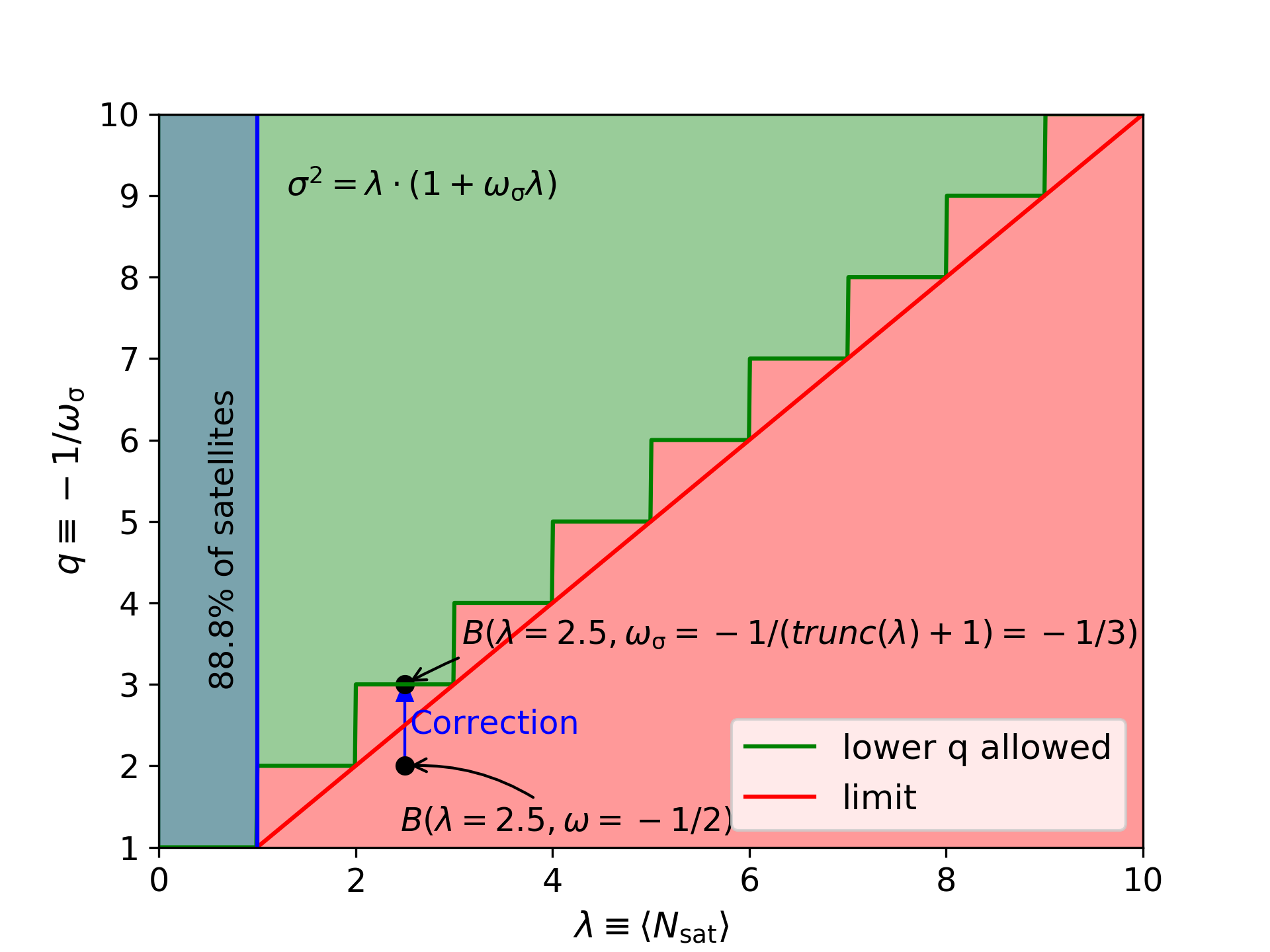}
    \caption{The parameter space ($\lambda,q \equiv -1/\omega_{\rm \sigma}$) of the binomial distribution. Considering the default HOD model, an 88.8 per cent of satellites reside in the blue region, in which the binomial distribution does not give unphysical negative variances. The points leading to unphysical negative variances, cover the red region below the red line, including it. The red region above the red line is out of the parameter space, since $q$ is an integer number. We use the ladder $q = -1/\omega_{\rm \sigma} = \rm{trunc} \left(\langle N_{\rm sat} \rangle\right)+1$ as the preferred boundary for corrections, ensuring: 1) the minimum effective variance to be positive and 2) $\langle N_{\rm sat} \rangle$ to be recovered. We show an example correcting a point in the red region of the parameter space: $\lambda = 2.5$, $\omega= -\frac{1}{2}$ to the nearest point in the green region of the parameter space with the same mean $\langle N_{\rm sat} \rangle$: $\lambda = 2.5$, $\omega_{\rm \sigma} = -\frac{1}{3}$.
    }
    \label{fig:range_binomial}
\end{figure}

For those haloes that contain, on average, one or more satellite galaxies, if $\omega < -1/\lambda$ (we remind that $\omega$ is related to the variance) the PDF becomes negative and the expression of the binomial variance described by \autoref{eqn: std_binomial} will give negative values of the variance, $\sigma^2 < 0$ (red area in \autoref{fig:range_binomial}).

To avoid unphysical cases with negative probabilities the HOD model must correct the input value of $\omega$ for those haloes, enhancing the variance only what is strictly necessary:

\begin{equation}
\label{eqn: serrated}
    \omega_{\rm \sigma} (M)= \left\lbrace\begin{array}{c}  -\frac{1}{\rm{trunc}(\langle N_{\rm sat} (M)\rangle)+1} \hspace{2mm} \rm{if} \hspace{2mm} \omega < -\frac{1}{\rm{trunc}\langle N_{\rm sat} (M)\rangle +1} \\  \omega \hspace{7mm} \rm{otherwise} \hspace{2mm} \end{array}\right.
\end{equation}
in which $\omega$ is the input parameter for the HOD model used for the entire catalogue, and $\omega_{\rm \sigma}$ is the value that will be finally used for a given halo mass $M$. We will use by default $\omega_{\rm \sigma}$ throughout this paper, except when it is an input parameter on the code. In this way, it is guaranteed that we always recover the correct mean number of satellites and a positive variance when we populate haloes with satellites.

We represent in \autoref{fig:HOD_scatter} the mean number of satellites $\lambda = \lambda(M)$ in a black solid line. We also focus on the orange solid lines, which represent the $\lambda \pm 1\sigma$ contours, in which $\sigma$ follows \autoref{eqn: std_binomial} for $\omega = -0.5$. Those lines are computed using 20 galaxy catalogue realizations. 

In \autoref{fig:range_binomial}, we represent the parameter space $\left\lbrace q = \frac{1}{\omega_{\rm \sigma}},\lambda \right\rbrace$. We show the limit when the PDF starts to be negative (red line), together with the limitation we have set in the parameter space (green staggered line). Above the green staggered line we do not need corrections in the parameter space and have $\omega_{\rm \sigma} = \omega$ (green area and blue area, the latter representing $\lambda < 1$). Below the red line we represent the region in which corrections have to be made: $\omega_{\rm \sigma} = -\frac{1}{\rm{trunc}(\langle N_{\rm sat} \rangle)+1}$ (red area). Despite green and red regions being filled continuously, $q$ is an integer. Therefore, the region between red and green lines is not populated by the binomial distribution.

\begin{figure}
    \centering
    \includegraphics[width=0.5\textwidth]{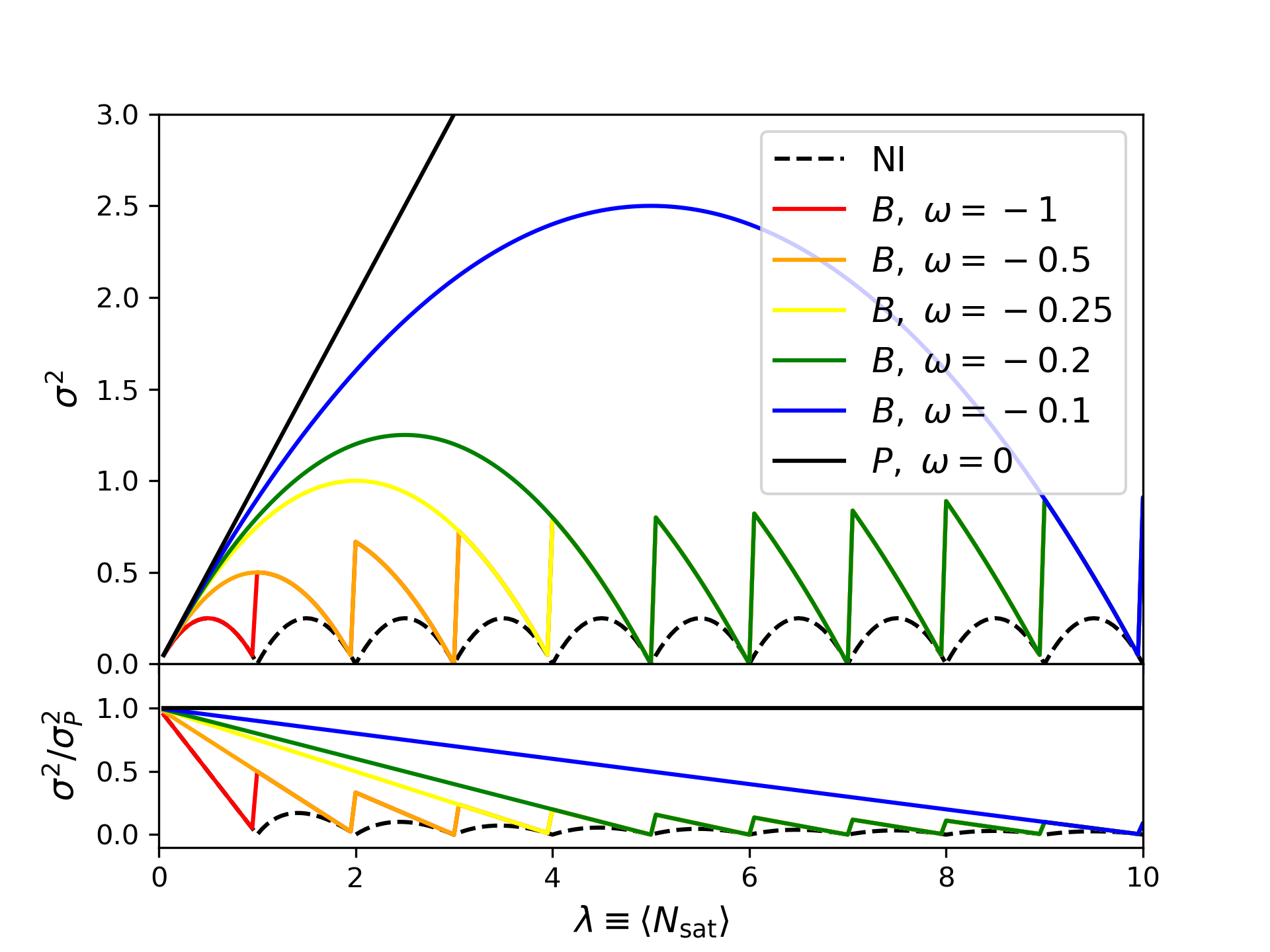}
    \caption{{\it Top}: Theoretical standard deviation $\sigma^2$ of the Nearest integer (NI), Poisson (P) and binomial (B) distributions as a function of the mean $\lambda$, for several input values of $\omega$. {\it Bottom}: ratio between all distributions and Poisson.}
    \label{fig:std_binomial}
\end{figure}

\autoref{fig:std_binomial} shows a set of binomial variances given by different input values of $\omega$ and mean number of satellite galaxies, $\lambda = \langle N_{\rm sat}\rangle$. They are compared with Poisson and Nearest Integer variances. As we can see, binomial variances are lower than the Poisson variance and higher than the Nearest Integer one: for low $\lambda$ we see the inverted arc shape which is the natural shape of \autoref{eqn: std_binomial}. For high $\lambda$ there is a serrated behaviour arising from the application of \autoref{eqn: serrated} to avoid the unphysical negative values of the variance. This serrated behaviour is directly related to the green ladder in \autoref{fig:range_binomial}.

\subsection{Extended Binomial distribution}
\label{sec:beyond_binomial}

As we described in \autoref{sec: binomial}, $\omega_{\rm \sigma}$ is limited to discrete values in the binomial distribution, since $\omega_{\rm \sigma} = -1/q$ and $q$ range is limited to natural numbers. Hence, the variance $\sigma^2$ for a particular $\langle N_{\rm sat}\rangle$, given by \autoref{eqn:variance_omega}, is also limited to discrete values. In this section our objective is to provide an extension to continuous values of $\omega_{\rm \sigma}$ allowing the variance to take continuous values for a given $\langle N_{\rm sat}\rangle$. For this purpose, we define a new probability distribution function, the extended binomial distribution $B_{\rm ext}$:

\begin{equation}
\label{eqn:extended_binomial}
    B_{\rm ext} \left(N_{\rm sat};\lambda,\omega_{\rm \sigma}\right)= f_{\rm q} \left(N_{\rm sat};\lambda,\omega_{\rm \sigma}\right)\cdot  B \left(N_{\rm sat};p,q\right) \hspace{2mm},
\end{equation}
where $B \left(N_{\rm sat};p,q\right)$ is the binomial distribution. $f_{\rm q} \left(N_{\rm sat};\lambda,\omega_{\rm \sigma}\right)$ is introduced to extend our range of sub-Poissonian variances since our input parameter $\omega_{\rm \sigma}$ can now take continuous values: $-1 \leq \omega_{\rm \sigma} < 0$. We have that $q \equiv \rm{ceil}\left(-1/\omega_{\rm \sigma}\right)$, in which $\rm{ceil}(x)$ is the closest upper integer to $x$ (we justify in appendix \ref{ap: relation_q_omega} this relation between $q$ and $\omega_{\rm \sigma}$). Note that $q \in \mathbf{N}$ as required. $f_{\rm q} \left(N_{\rm sat};\lambda,\omega_{\rm \sigma}\right)$ is also computed such that $B_{\rm ext}$ matches the first three binomial moments (see the detailed calculations in appendix \ref{ap: f_calculation}).

To evaluate $f_{\rm q} \left(N_{\rm sat};\lambda,\omega_{\rm \sigma}\right)$, $\lambda$ is evaluated using \autoref{eqn: Nsat_mean} as usual, $\omega_{\rm \sigma}$ is obtained from the input value of $\omega$, applying the correction introduced in \autoref{sec:correctionbinomial} and $q \equiv \rm{ceil}\left(-1/\omega_{\rm \sigma}\right)$.  Finally, to evaluate $B \left(N_{\rm sat};p,q\right)$, we have again that $q \equiv \rm{ceil}\left(-1/\omega_{\rm \sigma}\right)$ and also $p = \lambda/q$.

The calculation of the analytic expression of $f_{\rm q} \left(N_{\rm sat};\lambda,\omega_{\rm \sigma}\right)$ is detailed in appendix \ref{ap: f_calculation}. $f_{\rm q} \left(N_{\rm sat};\lambda,\omega_{\rm \sigma}\right)$ has to be calculated separately for each $q$ solving $q+1$ equations: a normalization equation (such that the PDF sums to unity), an equation for the mean number of satellites $\lambda$ and $q-1$ equations for the subsequent central moments. We solve for $q=1$ (trivial case, $\omega_{\rm \sigma} = -1$), $q=2$ $\left(\omega_{\rm \sigma} \in \left(-1,-\frac{1}{2}\right]\right)$ and $q=3$ $\left(\omega_{\rm \sigma} \in \left(-\frac{1}{2},-\frac{1}{3}\right] \right)$, by solving the corresponding equations.
For $q \geq 2$, we can choose the variance to have the expression of \autoref{eqn:variance_omega} \footnote{The expression of the variance for $q=1$, which cannot be imposed, coincides with \autoref{eqn:variance_omega}, see appendix \ref{ap: f_calculation}}. For $q \geq 3$, we choose the skewness to match the expression given by \autoref{eqn:skewness} \footnote{The expression of the skewness for $q = 1$ and $q = 2$ cannot be imposed, and for $q = 2$, it does not coincide with \autoref{eqn:skewness} (see gray region of \autoref{fig:Bext_Generalcase} and appendix \ref{ap: f_calculation})}.
For $q \geq 4$, one would need to specify the following higher order central moments, becoming increasingly complicated. In appendix \ref{ap: f_calculation} we propose a generalisation of our $q=2$ and $q=3$ solutions for $q \geq 4$. We find this generalisation to work well in most of the parameter space, but we also find a small part of the parameter space in which negative probabilities arise ($f_q < 0$). This has a very small impact in our parameterisation as we will see in \autoref{sec:performance}.

Note that the binomial distribution is a particular case of $B_{\rm ext}$, with $\omega_{\rm \sigma} = -1/q$. (See \autoref{sec: binomial}). In those cases $f_{\rm q} = 1$.

\subsection{Mitigating numerical limitations of the satellite PDF} 
\label{sec: gamma_limitation}

$\Gamma$ functions are used in the negative binomial distribution, \autoref{eqn:NegBinomial}, the binomial distribution, \autoref{eqn:binomial}, and its extension, \autoref{eqn:extended_binomial}. Numerical errors for these $\Gamma$ functions may arise for very large arguments: for $\Gamma (y \geq y_{\rm \Gamma lim})$ an overflow is produced.

Then, the implementation of the probability distribution function fails in the task to place satellite galaxies in the haloes, obtaining a galaxy catalogue without satellites, thus decreasing the input number density of galaxies.  $\Gamma (y \geq y_{\rm \Gamma lim})$ imposes a limitation on the parameter space $(q,N_{\rm sat})$: 

\begin{equation}
\label{eqn: limitation}
    y = q + g(N_{\rm sat}) < y_{\rm{max}} \, ,
\end{equation}
where $g(N_{\rm sat})$ is the remaining $\Gamma$ argument.

In this work, we compute $\Gamma$ functions using \textsc{tgamma} from the $\sc{math.h}$ library in \textsc{C}. With this particularities, $y_{\rm{max}} = 171.7$. 
Since $q$ is inversely proportional to $\omega_{\rm \sigma}$, \autoref{eqn: limitation} sets a lower limit in $\omega_{\rm \sigma}$.  

As the $\Gamma$ functions enter in both \autoref{eqn:NegBinomial} and \autoref{eqn:binomial} as a division, we define the following function:

\begin{equation}
\label{eqn:Gm}
Gm (N_{\rm sat},q) \equiv \frac{\Gamma(q+h(N_{\rm sat}))}{\Gamma(q + g(N_{\rm sat}))} \hspace{2mm} ; \hspace{2mm} h(N_{\rm sat}) \geq g(N_{\rm sat}) \hspace{2mm}.
\end{equation}

The above function fails for $\Gamma (y \geq y_{\rm \Gamma lim})$ when \autoref{eqn: limitation} is not satisfied due to numerical error. To avoid this we propose to use the product function, which is mathematically equivalent to \autoref{eqn:Gm} and does not fail when \autoref{eqn: limitation} is not satisfied:

\begin{equation}
\label{eqn:Pr}
    Pr(N_{\rm sat},q) \equiv \prod_{i = 0}^{h(N_{\rm sat})-g(N_{\rm sat})} \left(i+q + g(N_{\rm sat})\right) = Gm (N_{\rm sat},q)
\end{equation}

With this substitution, we can rewrite the binomial and negative binomial with products as follows:

\begin{equation}
\label{eqn:NegBinomial*}
    \rm{NB_{\Pi}} (N_{\rm sat}; p,q) = \frac{\prod_{i=0}^{N_{\rm sat}} \left(i+q\right)}{\Gamma(N_{\rm sat}+1)} p^q \left(1-p\right)^{N_{\rm sat}} \, ,
\end{equation}

\begin{equation}
\label{eqn:binomial*}
     \rm{B_{\Pi}} (N_{\rm sat}; q,p) = \frac{\prod_{i=0}^{N_{\rm sat}} \left(i+q+1-N_{\rm sat}\right)}{\Gamma \left(N_{\rm sat}+1\right)} p^{N_{\rm sat}} \left(1-p\right)^{q-N_{\rm sat}} \, .
\end{equation}

Except for the discussion introduced in this section, throughout this work we will always use \autoref{eqn:NegBinomial*} and \autoref{eqn:binomial*} when Negative binomial and binomial functions are needed, respectively. In this section, they are denoted as $\rm{NB_{\Pi}}$ and $\rm{B_{\Pi}}$, but in further sections we will use the notation $\rm{NB}$ and $\rm{B}$ for simplicity.

In our model, this correction is only applicable for both the negative binomial distribution and for the binomial distribution. The extended binomial distribution also suffers from $\Gamma$ overflows on the $f_q$ function, but it cannot be corrected, since this correction is only applicable when two $\Gamma(q + g(N_{\rm sat}))$ functions stay at both sides of a division. This is not the case of $B_{\rm ext}$ (See \autoref{eqn:extended_binomial}, \autoref{eq:f} and \autoref{eq:gn}). 

We now quantify the effect of this change on the PDFs. When $\omega_{\rm \sigma} > 0$ we use the negative binomial distribution and when $\omega_{\rm \sigma} < 0$, we use the binomial distribution.

\begin{figure}
    \centering
    \includegraphics[width=0.5\textwidth]{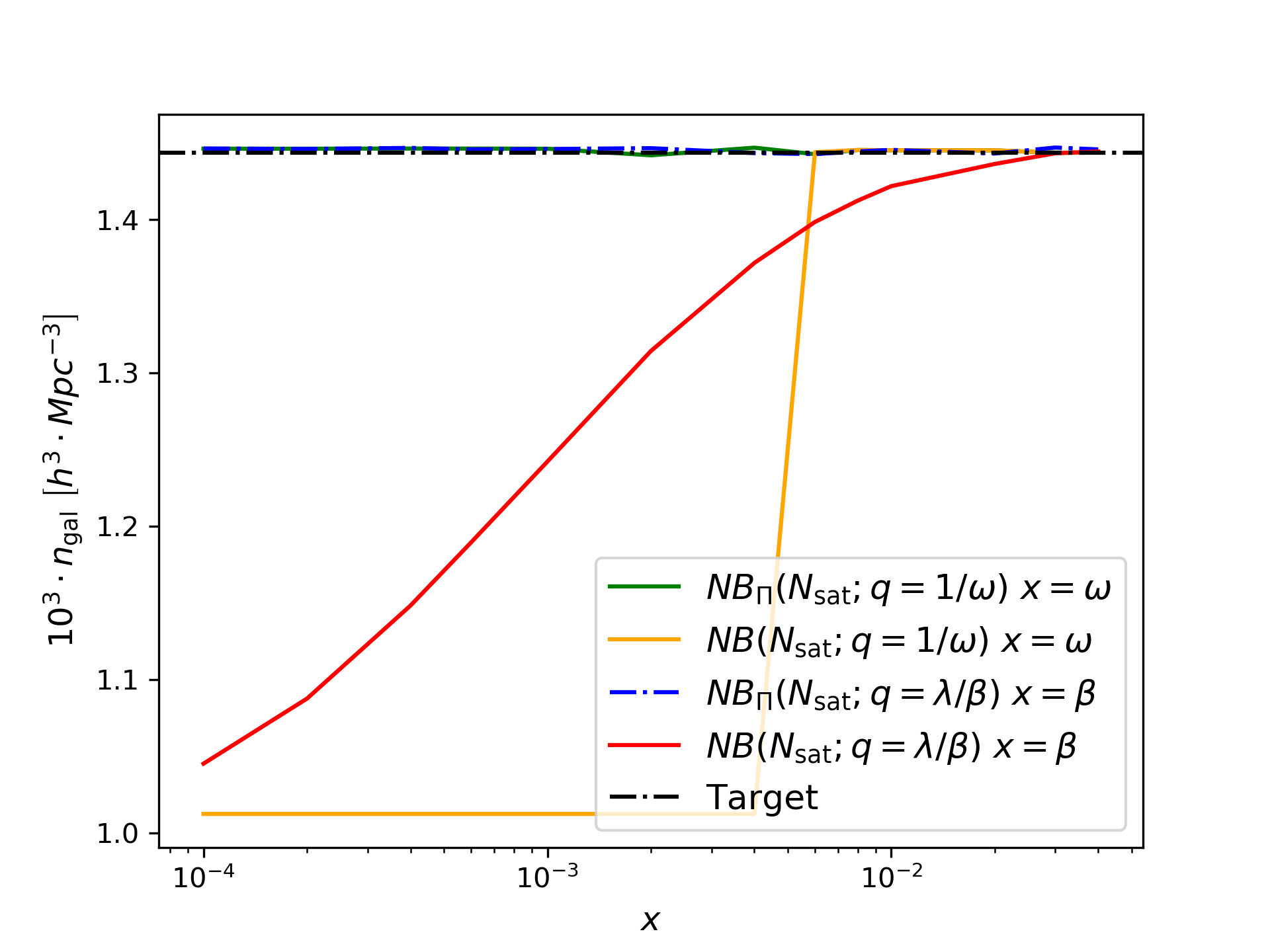}
    \caption{Ratio between the recovered and the target number density multiplied by $10^3$ for clarity, as a function of $x > 0$. The orange line is obtained using $\rm{NB} (N_{\rm sat};q=1/\omega)$ \autoref{eqn:NegBinomial}. When $\Gamma (y > y_{\Gamma lim})$ in \autoref{eqn:NegBinomial}, numerical computation of this function fails and no satellite galaxies are assigned to dark matter haloes, and thus the recovered number density decreases abruptly. The red line is obtained considering $\rm{NB} (N_{\rm sat} ;q=\lambda/\beta)$, which has a smoother decay when $\beta \xrightarrow[]{} 0$. Green and blue lines are obtained using $\rm{NB}_{\Pi}(N_{\rm sat};q = 1/\omega)$ and $\rm{NB}_{\Pi}(N_{\rm sat};q=\lambda/\beta)$, respectively. For both cases we recover correctly the number density in the whole range of $\omega$. Similar results are found for $\omega < 0$ using $\rm{B} (N_{\rm sat};q = -1/\omega)$, compared to $\rm{B}_{\Pi}(N_{\rm sat};q = -1/\omega)$. We use the default HOD model to compute the galaxy catalogues (\autoref{tab: default_HOD}). }
    \label{fig:ngal_fsat_gamma}.
\end{figure}

We present in \autoref{fig:ngal_fsat_gamma} a comparison between the input number density $n_{\rm gal}$ and the HOD model output for this quantity, as a function of $\omega$ \footnote{Note that we used $\omega$ instead of $\omega_{\rm \sigma}$ since we are talking about an input parameter on the code}. We have computed model galaxy catalogues with $f_{\rm sat,input} = 0.3$, $n_{\rm gal,input} = 6n_{\rm eBOSS} = 1.44 \cdot 10^{-3}$ $\left[\emph{h}/\rm{Mpc}\right]^3$ and $\omega \in  \left[10^{-4},4\cdot 10^{-2}\right]$.

\autoref{fig:ngal_fsat_gamma} shows the negative binomial (\autoref{eqn:NegBinomial*}, green solid line) and the negative binomial using only $\Gamma$ functions (\autoref{eqn:NegBinomial}, orange solid line). In the latter case, since numerical errors imply losing satellite galaxies, we do not recover the input number density. We observe an abrupt decaying of $n_{\rm gal}$ at $\omega = 0.006$. A similar behaviour is found for the binomial distribution using products (\autoref{eqn:binomial*}) and only $\Gamma$ functions (\autoref{eqn:binomial}), respectively.

In this work, for the negative binomial distribution we use the definition  $q=1/\omega_{\rm \sigma}$, however in the literature $q$ has been also defined as $q = \langle N_{\rm sat} \rangle / \beta =1/\omega_{\rm \sigma} $ \citep{Avila_2020,Jimenez_2019}. In the first case we use $\omega_{\rm \sigma}$ to control the variance of the PDF and in the second case they use $\beta$ for the same purpose. The variance can be expressed as $\sigma^2 = \lambda \left(1 + \beta\right)$. We represent in \autoref{fig:ngal_fsat_gamma} the same comparison between the input number density and the HOD output, as a function of $\beta$. We observe similar results as before: the use of negative binomial with products implies recovering the number density in the whole range (See blue dashed line), while in the case of using only $\Gamma$ functions we observe a decay, which is smoother since $q$ depends on  $\langle N_{\rm sat} \rangle \propto M$.

\section{Computational implementation of the satellite PDF} 
\label{sec:performance}

Here we describe the algorithm we follow to choose a PDF for satellite galaxies given a target variance. This is quantified by an input parameter, $\omega$, to our halo occupation distribution (HOD) model. This free parameter quantifies how far the variance of the satellite PDF is from that for a Poisson distribution. Given any input $\omega$, unphysical negative probabilities can arise for some haloes containing more than one satellite galaxies, when $\omega < -1/ \left(\rm{trunc}(\langle N_{\rm sat} \rangle ) +1\right)$ (\autoref{sec:correctionbinomial}). To prevent this from happening, we modify the input parameter $\omega$ when needed, following \autoref{eqn: serrated}. We let denote $\omega_{\rm \sigma}$ the input parameter with the possible correction.

Depending on the value of $\omega_{\rm \sigma}$, the code for the HOD model will use different PDFs for satellite galaxies:

\begin{itemize}
    \item If $\omega_{\rm \sigma} = 0$ we assume a Poisson PDF (\autoref{eqn:Poisson}), with variance
    $\sigma^2=\lambda\equiv \langle N_{\rm sat} \rangle$.
    \item If $\omega_{\rm \sigma} > 0$ we assume a (corrected) negative binomial PDF (\autoref{eqn:NegBinomial*}) with super-Poisson variance, $\sigma^2 = \lambda \left(1+\omega_{\rm \sigma} \lambda\right)$ (\autoref{eqn:variance_omega}).
    \item If $\omega_{\rm \sigma} < 0$ we have two options for the satellite PDF, with a sub-Poisson variance:
    \begin{itemize}
         \item If $\omega_{\rm \sigma} < -1$ we assume a nearest integer PDF (\autoref{eqn: NI}), which has the smallest possible variance.
         \item Otherwise we assume the $B_{\rm sub-P}$ PDF (\autoref{sec: PDFextensions}) with a continuous sub-Poisson variance, $\sigma^2 \equiv \lambda \left(1+\omega_{\rm \sigma} \lambda\right)$ (\autoref{eqn: std_binomial}).
  \end{itemize}
\end{itemize}

The $B_{\rm sub-P}$ PDF is based on the (extended) binomial distributions introduced in \autoref{sec: PDFextensions}. By default, we use the new PDF that we have introduced in \autoref{sec:beyond_binomial} to allow for continuous values of $\omega_{\rm \sigma}$. This new PDF is the extended binomial, $B_{\rm ext}=f_{\rm q}B$ (\autoref{eqn:extended_binomial}). Note that $B_{\rm ext}$, is  reduced to the binomial one, $B$, when $f_{\rm q}=1$. This occurs when $\omega_{\rm \sigma} = 1/q$, $\forall q\in \mathbb{N} ^{+}$.

We do not use the $B_{\rm ext}$ for the $B_{\rm sub-P}$ PDF, in one case. When $\omega_{\rm \sigma} > -1/y_{\rm \Gamma lim}$ we use $B (q = \rm{ceil}\left(-1/\omega_{\rm \sigma}\right))$ (\autoref{eqn:binomial*}). For this range of $\omega_{\rm \sigma}$, $f_{\rm q}$ cannot be evaluated  due to the numerical limitations of using $\Gamma$ functions (see \autoref{sec: gamma_limitation} and \autoref{ap: f_calculation}). And thus, $B_{\rm ext}=f_{\rm q}B$ cannot be computed in this range.

Below, we evaluate the numerical performance of $B_{\rm sub-P}$.

\subsection{Performance for a uniform distribution}
\label{sec:general_case}

\begin{figure*}
    \centering
    \includegraphics[trim=40 60 20 60, clip, width=0.85\linewidth]{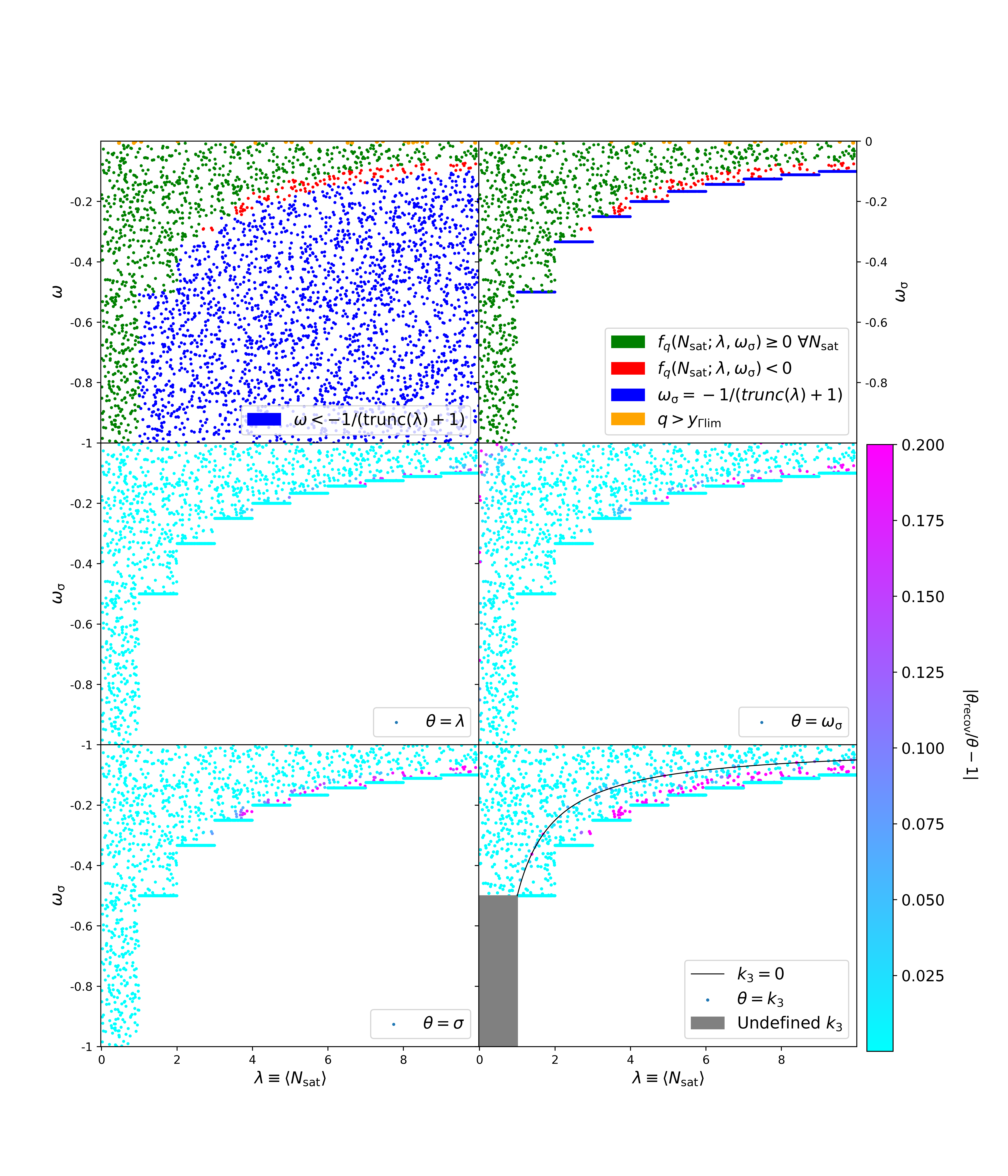}
    \caption{{\it Top left}: $3500$ random points with $\omega \in \left[-1,0\right)$ and $\lambda \in \left(0,10\right)$. We distinguish green points: $f_{\rm q}(N_{\rm sat};\lambda,\omega)> 0 $ $\forall N_{\rm sat}$, red points, in which $f_{\rm q}(N_{\rm sat};\lambda,\omega)< 0 $ for some $N_{\rm sat}$. We also consider points with $\omega < -1/(\rm{trunc}(\lambda)+1)$ and finally those with $q > y_{\rm \Gamma lim}$, represented in orange. {\it Top right}: correcting $\omega$ in favor of $\omega_{\rm \sigma}$ for blue points.   {\it Middle left}: we compare the mean recovered $\lambda_{\rm recov}$ with its respective fiducial value $\lambda$ for all points.  {\it Middle right}: we make the same comparison for $\omega_{\rm \sigma}$.  {\it Lower left}: same for $\sigma$.  {\it Lower right}: we do the same analysis for the skewness $k_3$. We represent $3290$ points, since we exclude points placed in the grey area and we also represent $k_3 = 0$.}
    \label{fig:Bext_Generalcase}
\end{figure*}

We evaluate the $B_{\rm sub-P}$ implementation introduced in \autoref{sec: PDFextensions} as a mathematical tool that can be applied when a continuous range of sub-Poisson variances are needed. For this purpose, we generate $3500$ points distributed randomly with $\omega \in \left[-1,0\right)$ and $\lambda = \left[0,10\right]$ (this range is motivated in \autoref{sec: par_space_galaxy_catalogues}). This uniform distribution is shown in the top left panel of \autoref{fig:Bext_Generalcase}. Such number of pairs $\left\lbrace\omega,\lambda \right\rbrace$ can provide a reliable estimation of how $B_{\rm sub-P}$ behaves when $f_{\rm q} < 0$ (red points in the top left of \autoref{fig:Bext_Generalcase}).

In the upper-left plot of \autoref{fig:Bext_Generalcase} we distinguish between four types of points, representing the three cases described above for $B_{\rm sub-P}$, that is, the use of $B_{\rm ext}$, green and red points in the top panels of \autoref{fig:Bext_Generalcase}; $B (q = \rm{trunc}\left(-1/\omega_{\rm \sigma}\right))$, orange points; and $B (\omega_{\rm \sigma})$, dark blue points. The red points in \autoref{fig:Bext_Generalcase}, correspond to negative probabilities due to $f_{\rm q} \left(N_{\rm sat};\lambda,\omega_{\rm \sigma}\right) < 0$. Points with $f_{\rm q} < 0$, represent potentially problematic regions in the parameter space, where errors could arise when recovering the input mean, variance and higher-order moments of the distribution. This region is centered around ($\lambda >1,\omega_{\rm \sigma} > -1/4$). We evaluate the impact of these below.

In the upper panels of \autoref{fig:Bext_Generalcase} we show the effect of correcting the input $\omega$, dark blue points uniformly distributed on the left, when $\omega < -1/ \left(\rm{trunc}(\langle N_{\rm sat} \rangle ) +1\right)$ (\autoref{sec:correctionbinomial}). The resulting $\omega_{\rm \sigma}$ parameter has a serrated behaviour, as shown by the dark blue points in the upper-right panel of \autoref{fig:Bext_Generalcase}.

We address the errors introduced by the corrections mentioned above by studying $10^6$ realisations of $B_{\rm sub-P}$ for each pair $(\omega_{\rm \sigma},\lambda)$ shown in the top right panel of \autoref{fig:Bext_Generalcase}. We compute the first three central moments: the mean $\lambda_{\rm recov}$, the variance $\sigma_{\rm recov}$ and the skewness $k_{3,\rm recov}$ of those realisations (for the skewness we only consider $\omega>-0.5$ since this is the range of applicability of \autoref{eqn:skewness}). We also compute $\omega_{\rm \sigma, recov}$ from ${\lambda_{\rm recov},\sigma_{\rm recov}}$ inverting \autoref{eqn:variance_omega}. Finally, we compare the results with the inputs $\lambda$, $\omega_{\rm \sigma}$, $\sigma$ and $k_3$.    

In the middle and lower panels of \autoref{fig:Bext_Generalcase}, we represent for each one of the points the following quantity: $|\theta_{\rm recov} /\theta-1|$, for $\theta = \lambda,\omega_{\rm \sigma},\sigma,k_3$, which represents the differences between recovered and input moments. As expected, high values of $|\theta_{\rm recov} /\theta-1|$ are found in the region containing points with $f_{\rm q}(N_{\rm sat};\lambda,\omega_{\rm \sigma})< 0 $, in which $\lambda > 2$ and $\omega_{\rm \sigma} > -1/3$. We get the largest errors when we consider the skewness $k_3$ and the lowest errors when we consider the mean $\lambda$. Errors arise also in the $\theta = \omega_{\rm \sigma}$ subplot for $(\lambda,\omega_{\rm \sigma} \xrightarrow{} 0)$: 
This is found to be numerical noise, and the relative errors get smaller as we increase the number of realisations. 
Finally, we also see deviations in the $\theta = k_3$ subplot, near to the $k_3 = 0$ curve, again simply due to statistical limitations. To sum up, in those two last cases there is not an intrinsic bias, since errors simply decrease when we consider a larger number of realizations. Intrinsic bias in the computation of moments arise in the region of the parameter space corresponding to $f_{\rm q}(N_{\rm sat};\lambda,\omega_{\rm \sigma})< 0 $.

\begin{figure}
    \centering
    \includegraphics[width=1.05\linewidth]{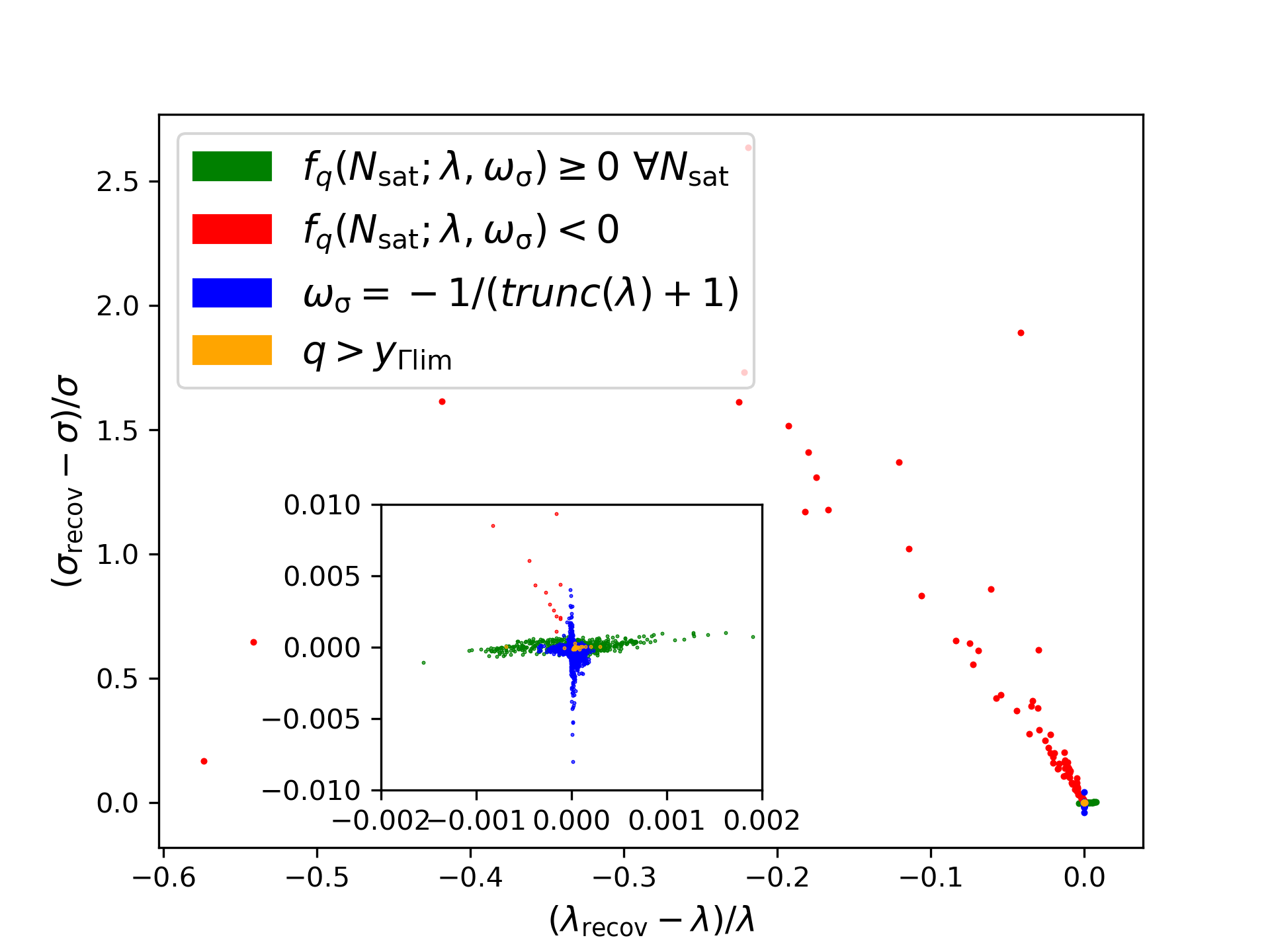}
    \caption{Difference between the recovered and input values of $\sigma$ and $\lambda$, divided by its respective input values. We represent $3500$ points, divided in four flags as in the upper plots of \autoref{fig:Bext_Generalcase}. The most relevant errors which introduce a real bias in the results are found for $f_{\rm q}(N_{\rm sat};\lambda,\omega_{\rm \sigma}) < 0$. The errors for points corresponding to the other flags are best shown in the zoom panel. Those are smaller and due to the limited number of realizations taken into account to compute $\lambda$ and $\sigma$.}
    \label{fig:errors_position}
\end{figure}

Since our main interest is to recover correctly the mean number of satellites and the variance, in \autoref{fig:errors_position} we show the differences between recovered and input values of $\lambda$ and $\sigma$. As we can see, only red points ($f_{\rm q} < 0$) have appreciable errors in our analysis. As we can see for those points, less of them ends up with a higher error values and higher discrepancies between $\lambda$ and $\lambda_{\rm recov}$ involve also higher discrepancies between $\sigma$ and $\sigma_{\rm recov}$ (Note that both parameters are related by \autoref{eqn:variance_omega}). In addition, we find $\lambda_{\rm recov} < \lambda$ while $\sigma_{\rm recov} > \sigma$. 

We discussed so far where we can find in our parameter space differences between recovered and input values of our parameters $\lambda, \omega_{\rm \sigma},\sigma$ and $k_3$. Now we can focus on determining how frequent those errors are. Considering $\theta = \lambda (\theta =\sigma)$, a $2$ ($2.8$) per cent of the points in the parameters range uniformly explored in \autoref{fig:Bext_Generalcase}  have errors greater than $0.5$ per cent. In contrast, if we consider $\theta = k_3$, a $9.5$ per cent of the points have errors greater than $0.5$ per cent.  

\subsection{Performance for galaxy catalogues}
\label{sec: par_space_galaxy_catalogues}

In the previous subsection we evaluated how $B_{\rm sub-P}$ performed in a uniform $\left\lbrace\omega,\lambda \right\rbrace$ space. Now, we evaluate the performance of $B_{\rm sub-P}$ for galaxy catalogue generation considering a reduced parameter space to isolate points in which $f_q < 0$. First, let us recall that we are considering a growing power law for $\langle N_{\rm sat} (M)\rangle$ (\autoref{eqn: Nsat_mean}, \autoref{fig:HOD_scatter}). At the same time, the halo mass function is a decreasing power law: we have much less massive haloes than lighter ones. This translates to a lesser number of haloes with a higher average number of satellites. If we consider \autoref{fig:Bext_Generalcase}, the difference we make is to consider progressively less number of points with higher $\langle N_{\rm sat}\rangle$.

We consider at first instance 1,000,000 points in the following parameter space: $\left\lbrace\omega \in \left(-\frac{1}{3},0\right),\lambda = \langle N_{\rm sat}(M) \rangle \right\rbrace$, in which $\lambda$ is obtained from $1,000,000$ \unit \footnote{We expect almost no variation in the result if \orsim halo masses are used, since both simulations share similar halo mass functions.} halo masses selected randomly applying \autoref{eqn: Nsat_mean} and considering $f_{\rm sat} = 0.65$ \footnote{For a particular halo mass, when the fraction of satellites is higher, $\lambda$ is also higher. Since the effect we want to evaluate arises for high values of $\lambda$ we are considering a high value of the fraction of satellites as a worst-case scenario.}. All the evaluated points stay at $0 < \lambda < 10$, motivating the limits imposed in the parameter space considered in \autoref{sec:general_case} \footnote{Although we do not use substructure here, it is worth noting, that \unit most massive haloes can contain over 100 substructures.} .
Note that we skipped $\omega \in \left(-1,-\frac{1}{3}\right)$, since in those cases $f_{\rm q} \geq 0$ $\forall N_{\rm sat}$. We motivate below further cuts to isolate the problematic region ($f_q < 0$) found in the last subsection, to avoid a large number of unnecessary $B_{\rm sub-P}$ evaluations.

We exclude points with $\lambda \leq 1$, since it can be demonstrated mathematically that $f_{\rm q}\left(N_{\rm sat};\lambda \leq 1,\omega\right)) \geq 0$. This is also demonstrated numerically in our previous test (\autoref{sec:general_case}). Finally, we also exclude points with $\omega <- 1/(\rm{trunc}(\lambda)+1)$ (equivalent to blue points in \autoref{fig:Bext_Generalcase}). For those points we remind that \autoref{eqn: serrated} is applied and the input value of $\omega$ is substituted by $\omega_{\rm \sigma} = -1/(\rm{trunc}(\lambda)+1)$. All values of $\omega_{\rm \sigma} = -1/(\rm{trunc}(\lambda)+1)$ fall in the range of the binomial distribution, and then we have that $f_q = 1$ for all those points (that is, $B_{\rm ext} = B$). We remind that those selections we made are motivated to isolate the problematic region of the parameter space in which $f_q < 0$. We have that $851$ points (from the initial $1,000,000$ points) passed the selection. We focus on those points to determine errors in the central moments.

In particular we compute, as in \autoref{sec:general_case}, $10^6$ realizations of $B_{\rm sub-P} $ for all selected points. We use the results to compare the recovered and fiducial values of $\lambda$, $\omega_{\rm \sigma}$, $\sigma$ and $k_3$. In this test, we focus now on quantifying how frequent errors on $\theta = \lambda, \omega_{\rm \sigma},\sigma$ and $k_3$ are. In this particular test, we are interested only in the errors arising when $f_{\rm q} (N_{\rm sat};\lambda,\omega_{\rm \sigma})<0$. Thus, the statistics are obtained considering the number of points with errors greater than $|\theta_{\rm recov}/\theta -1|$ ($\theta = \lambda,\omega_{\rm \sigma},\sigma,k_3$) for the $851$ selected points, but we normalize the result adding zero errors for the remaning points in which we already knew that $f_q \geq 0$ (In total, $1,000,000$ points). Note that in order to obtain a reliable statistic for galaxy catalogue generation, we need to consider all the original points. The selection was done to compute the errors only in the isolated region of the parameter space in which  $f_q < 0$ was possible, thus avoiding a high computational cost.

\begin{figure}
    \centering
    \includegraphics[width=1.05\linewidth]{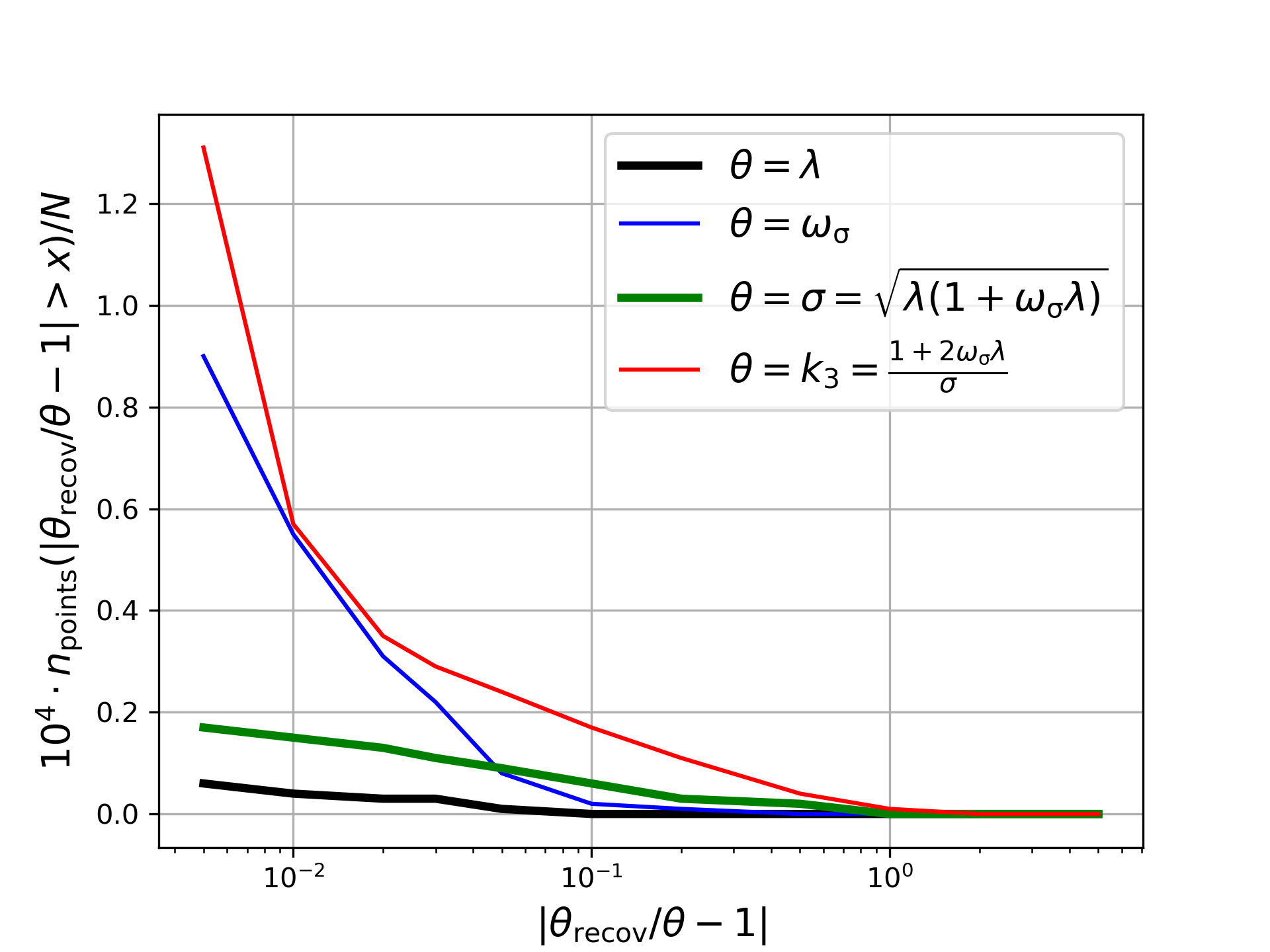}
    \caption{We represent the normalised number of $B_{\rm sub-P}$ realization errors (multiplied by $10^4$ due to axis clarity) higher or equal to $|\theta_{\rm recov}/\theta -1|$ divided by the total number of points considered, $N = 10^6$, with colours as indicated by the legend}
    \label{fig:Bext_mockcase_errors}
\end{figure}

In \autoref{fig:Bext_mockcase_errors} we present the number of $B_{\rm sub-P} $ realizations that have an error greater than the $|\theta_{\rm recov}/\theta -1|$ values on the x-axis: only 6 per million points have and error greater than $0.5 $ per cent recovering the mean. In the same way we find that only 17 per million points have an error greater than $0.5 $ per cent recovering $\sigma$ and only 131 per million points recovering the skewness. The discrepancies between the fiducial and recovered values arise at high values of $\lambda$, which correspond to very massive haloes, that are much less frequent in nature. To sum up, in the particular task of mock generation, numerical errors are not expected to have an impact in the resulting galaxy distribution of the catalogues.

\section{2PCF and the satellite PDF} 
\label{sec:clustering}

We expect the satellite PDF to affect the pair-counts of galaxies within a halo, i.e., the 2-point correlation (2PCF) 1-halo term. On the other hand at large scales (in the 2-halo term), the 2PCF would be unaffected by the PDF itself. This intuition is indeed confirmed when we measure the 2PCF HOD catalogues generated with different PDF variances.

\begin{figure}
    \centering
    \includegraphics[width=0.45\textwidth]{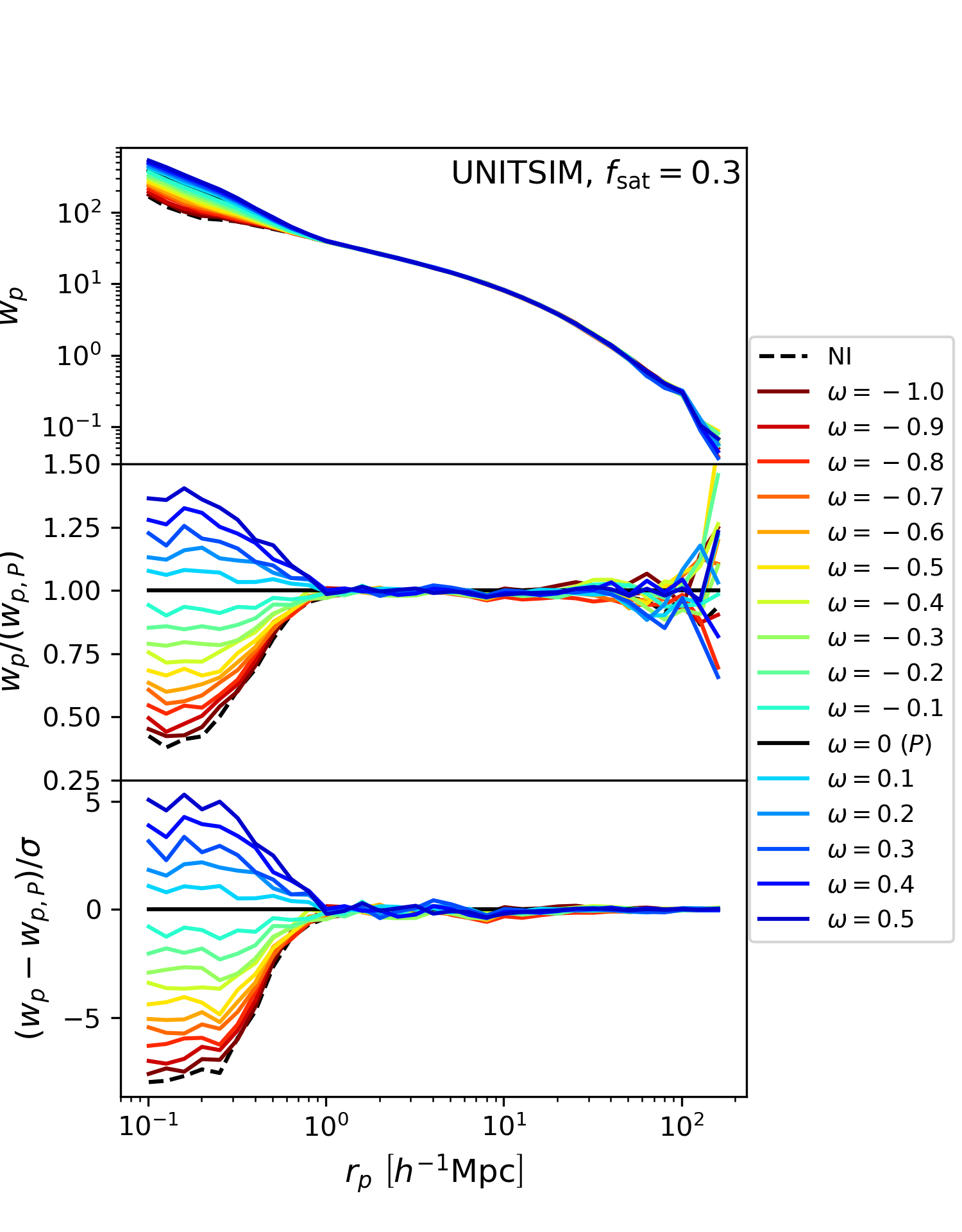}
    \caption{Projected correlation functions calculated from \unit simulations $w_{\rm p}$ considering a fraction of satellites $f_{\rm sat} = 0.3$. We consider Negative Binomial distributions ($\omega > 0$), Poisson distribution ($\omega = 0$), $B_{\rm sub-P}$ ($-1 \leq \omega < 0$) and Nearest Integer distribution.  {\it Top}: projected correlation function $w_{\rm p}$.  {\it Middle}: ratio between all $w_{\rm p}$ with Poisson $w_{\rm {p,P}}$ ($\omega = 0$).  {\it Bottom}: difference between all $w_{\rm p}$ and Poisson $w_{\rm {p,P}}$ divided by the error.
    }
    \label{fig:wp_all_PDF}
\end{figure}

In \autoref{fig:wp_all_PDF} we can see how the variance of the probability distribution used to populate haloes with satellite galaxies affects the projected correlation function $w_{\rm p}(r_{\rm p})$. 

In the upper panel we see the general shape of $w_{\rm p} (r_{\rm p})$. As we can see, it follows the typical decaying power law as a function of the projected distance. 

In the middle panel we divide all $w_{\rm p} (r_{\rm p})$ considered by the Poisson $w_{\rm p,P} (r_{\rm p})$. We can see that small scales are affected as expected: clustering rises when the variance considered is larger. A larger variance favor more instances with more than one satellite in the same halo, increasing precisely the 1-halo term. Since the linear bias has been fixed, intermediate and large scales are mostly not affected since galaxy pairs comes from different haloes and now former differences are averaged out. 

In the lower panel we can see the difference between all $w_{\rm p}$ and the Poisson $w_{\rm p,P}$ divided by the error calculated using jackknife resampling. We detail the calculation of jackknife errors in the following subsection. We show that variations in the one-halo term of the projected correlation function induced by changes in the PDF are clearly significant, showing the constraining power that $w_{\rm p} (r_{\rm p})$ has on the variance on small scales. 

We do not show it here but we find that the dependence of the monopole and quadrupole on the PDF variance is more modest, in line with the results from \citet{Avila_2020}. Since we use mostly the one-halo term of the projected correlation function when fitting galaxy catalogues with eBOSS data, we also use the linear scales of $\xi_0(s)$ and $\xi_2(s)$ in our fits. In this regime, the monopole depends more strongly on the linear bias, and there is no appreciable dependence in the quadrupole, which is more affected by the velocity distribution.

\subsection{Jackknife resampling}
\label{sec:jackknife}

To estimate the covariance of the two-point functions y = $(w_{\rm p},\xi_0,\xi_2)$  introduced in this section, and also for the Count in Cells estimator introduced in \autoref{sec:CIC} we use jackknife resampling. 

In the case of two-point functions using \unit simulation, we construct $N_{\rm box} = 1000$ copies of a catalogue of galaxies, with $n  = n_{\rm eBOSS}$ and ($\omega$, $f_{\rm sat}$) = (-0.8,0.4). To each copy we extract a different sub-box of $L_{\rm cell} = 100 $ $\emph{h}^{-1}\rm{Mpc}$. We compute $y_i = (w_{\rm p,i},\xi_{0,i}, \xi_{2,i})$ for each copy and we apply:

\begin{equation}
    C_{\rm Jack} = \frac{N_{\rm box}-1}{N_{\rm box} } \sum_{i=1}^{N_{\rm box}} \sum_{j=1}^{N_{\rm box}} \left(y_{i}-y\right) \cdot \left(y_{j}-y\right)
\end{equation}

We finally need to rescale the variance, comparing the volume of the simulation and the eBOSS ELG volume:

\begin{equation}
    C = C_{\rm Jack} \cdot \frac{V_{\rm box}}{V_{\rm eBOSS}}
\end{equation}

In the case of \unit simulation, $V_{\rm box} = 1$ $\emph{h}^{-1}\rm{Gpc}$. 

Nevertheless, when considering two-point function using \orsim simulation, we have that $V_{\rm box} = 3$ $\emph{h}^{-1}\rm{Gpc}$. In that case, we divide the latter box in 27 sub-boxes of $1$ $\emph{h}^{-1}\rm{Gpc}$, we follow the steps outlined above for each one of the sub-boxes and we finally average the errors.

Finally, we compute the jackknife covariance for Count-in-cells following the steps described above, but using $N_{\rm box} = 125$.

\subsection{Application to model eBOSS ELGs}
\label{sec: fitting_data}

As an application we fit the eBOSS ELG data using HOD models with a range of satellite PDFs. This is done as an exercise and thus, we will only vary 2 parameters: the fraction of satellites $f_{\rm sat}$ and the PDF variance by setting free $\omega$. We get independent fits for \unit and \orsim simulations. 

We consider the following PDFs: Nearest Integer (NI), $B_{\rm sub-P}$ with $\omega \in [-1, -0.05]$, Poisson ($\omega = 0$) and Negative binomial $\omega \in [0.05,0.5]$. We consider a step $\Delta \omega = 0.05$ for the $B_{\rm sub-P}$ and the NB PDFs. This parameter space is shared by the two simulations used in this work. 

We also vary the fraction of satellites. For \unit, we consider $f_{\rm sat} \in [0.05, 0.75]$ with $\Delta f_{\rm sat} = 0.05$. We exclude $f_{\rm sat} > 0.75$ since in this case $\mu < M_{\rm min} = 10^{10.42}$ $M_{\odot} \emph{h}^{-1}$, implying a model in which most of the central galaxies have to be placed at small-mass unresolved haloes. We point out that this issue is particular to the HOD considered in this work. Other HOD models may allow more freedom in the fraction of satellites range. We exclude as well $f_{\rm sat} = 0$, corresponding to a galaxy catalogue populated only with central galaxies. For \orsim, we consider $f_{\rm sat} \in [0.05, 0.6]$, since $f_{\rm sat} > 0.6$ is implying $\mu < M_{\rm min} = 10^{10.58}$ $M_{\odot} \emph{h}^{-1}$. All remaining parameters are set to their default values (See \autoref{tab: default_HOD}). To increase the signal-to-noise ratio, the galaxy catalogues constructed have a number density $n_{\rm gal} = 6n_{\rm eBOSS}$.

We compute the $\chi^2$ to compare the reference data catalogue with each one of our galaxy catalogues. The expression of the $\chi^2$ as follows:

\begin{equation}
\label{eqn: chi2data}
    \chi^2 \left(\theta\right) = \left(y_{\rm data} - y_{\rm sim} \left(\theta\right)\right)^T C^{-1} \left(y_{\rm data} - y_{\rm  sim} \left(\theta\right) \right)
\end{equation}

in which $y = (w_{\rm p},\xi_0,\xi_2)$ , $\theta = \left(f_{\rm sat},\omega\right)$ and $C$ is the covariance matrix computed in \autoref{sec:jackknife}.

Clustering information is computed following \citet{Avila_2020}: we compare the projected correlation function $w_{\rm p} (r_{\rm p})$ evaluated at $0.19 < r_{\rm p} (\emph{h}^{-1}\rm{Mpc}) < 4.5$, the monopole $\xi_0 (s)$ evaluated at $20 < s (\emph{h}^{-1}\rm{Mpc})< 45$ and the quadrupole $\xi_2 (s)$ evaluated at $10 < s (\emph{h}^{-1}\rm{Mpc})< 25$ (22 points in total). Therefore, we consider (22-2 = 20) degrees of freedom.

Finally, we find the best fit galaxy catalogue extracting $\left(f_{\rm sat},\omega\right)$ values that correspond to the minimum $\chi^2$, $\chi_{\rm min}^2$.

\begin{figure}
    \centering
    \includegraphics[width=0.48\textwidth]{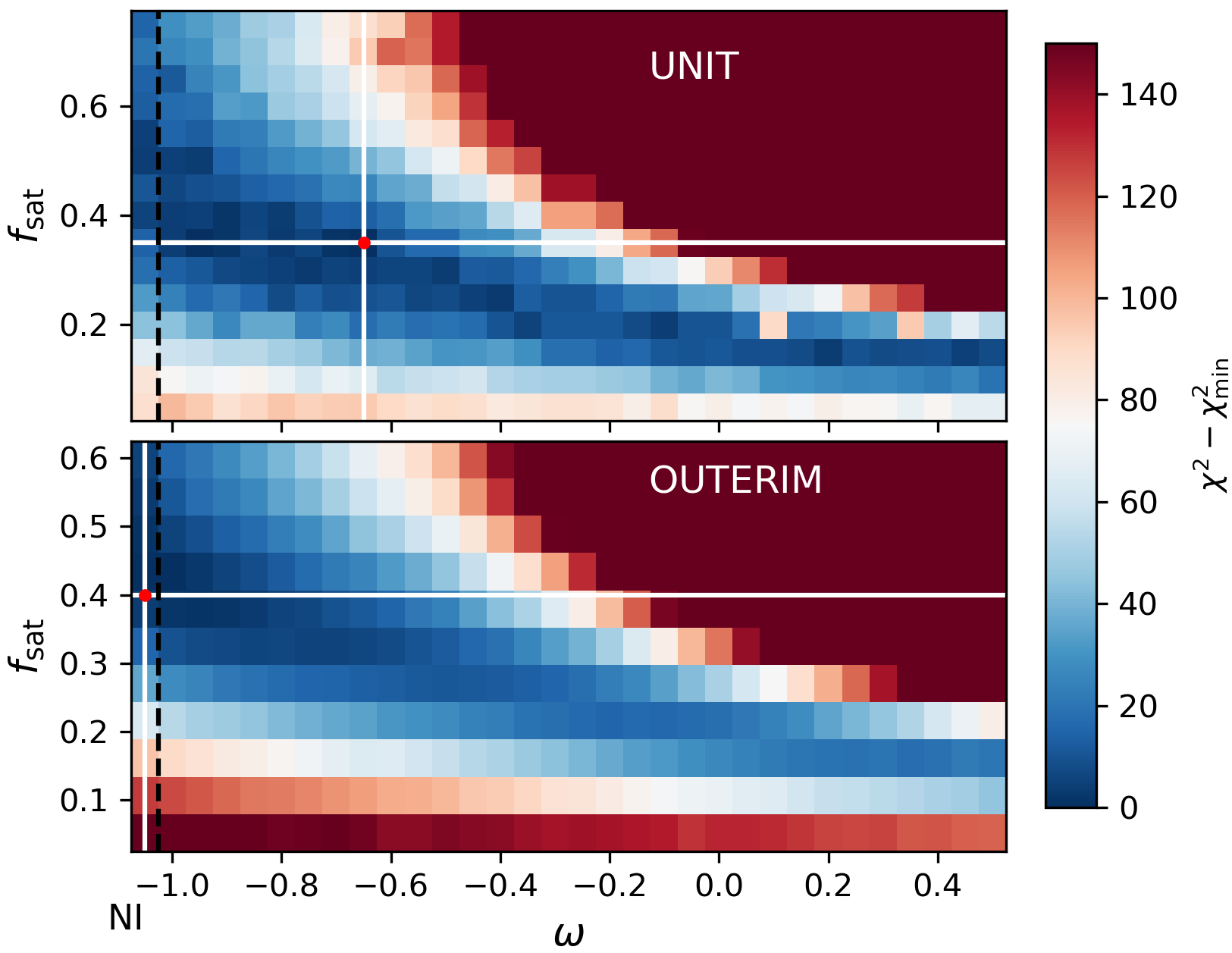}
    \caption{ {\it Top}: comparison between eBOSS ELG data and galaxy catalogues obtained form \unit simulation with $\omega$ and $f_{\rm sat}$ as free parameters. Specific ranges of the projected correlation function, the monopole and the quadrupole are used for the comparison. Each point represents the $\chi^2-\chi_{\rm min}^2$ for a particular point of ($\omega$,$f_{\rm sat}$) parameter space. $\chi^2-\chi_{\rm min}^2 = 0$ represented by a red point is the best fit to the data, corresponding to $(\omega , f_{\rm sat})= (-0.65, 0.35)$. Since $\omega < 0$, a sub-Poisson distribution is preferred, parametrised by the extended binomial distribution.  {\it Bottom}: same comparison using \orsim galaxy catalogues. The best fit prefers a Nearest Integer distribution and a fraction of satellites $f_{\rm sat}= 0.4$.}
    \label{fig:xi2_unit_orsim}
\end{figure}

In the top panel of  \autoref{fig:xi2_unit_orsim} we show a comparison between our \unit-based galaxy catalogues and eBOSS ELG data as our reference catalogue through the evaluation of $\chi^2 (f_{\rm sat},\omega) - \chi_{\rm min}^2$. As we can see, there is some degeneracy between the two parameters. If we increase the variance of the distribution $\omega$ or the fraction of satellites $f_{\rm sat}$ it also increases the one halo term of the projected correlation function. Both effects seem to compensate each other. Besides that, we find that the preferred galaxy catalogue has ($f_{\rm sat}$,$\omega$) = $(0.35,-0.65)$, corresponding to a catalogue in which the $35$ per cent of galaxies are satellites, and satellites are placed with a sub-Poisson variance, modelled with $B_{\rm ext}$ ($\omega = -0.65$). 

In the bottom panel of \autoref{fig:xi2_unit_orsim} we show $\chi^2 (f_{\rm sat},\omega) - \chi_{\rm min}^2$ for galaxy catalogues computed using \orsim simulation. The best fit prefers satellite galaxies to be distributed following a Nearest Integer distribution, with $40 $ per cent of the total number of galaxies as satellites ($f_{\rm sat} = 0.4$).

We can see that \orsim $\chi^2 (f_{\rm sat},\omega)$ function is smoother than the same quantity computed using \unit simulation, since the volume of the simulation is 27 times bigger (See \autoref{tab: sim_properties}). \unit fixed technique suppresses variance on large scales \citep{Chuang_2019}, but since small scales have an important weight in our $\chi^2$ fitting, it is not possible to benefit from this suppression. The results are not fully consistent between the two simulations. In particular, the difference of the best fits found for both simulations, considering \orsim simulation is around $ \sqrt{\Delta \chi^2} \approx 3.5 \sigma$. Nevertheless, this inconsistency is not surprising since both simulations are run with different cosmologies (See \autoref{tab: sim_properties}).

The results obtained are just a first test aiming to give a first use to the $B_{\rm sub-P}$ extension in a simple context of two free-parameter fitting ($f_{\rm sat}, \omega$) to eBOSS ELG data, using standard clustering measurements ($w_{\rm p} (r_{\rm p}),\xi_0(s),\xi_2(s)$), following \citet{Avila_2020}. 

We also want to see possible simulation-dependent differences in the results, as the use of different volumes and cosmologies on the $\chi^2$ fitting results. The actual best fit values are not so important in this work: here the focus is on the probability distribution function, and thus we fixed most HOD parameters.

\section{Count in Cells}
\label{sec:CIC}
In \citet{Avila_2020} the authors explored the HOD parameter space using clustering statistics, and degeneracies appeared between some of the parameters. We find similar degeneracies in this work in \autoref{fig:xi2_unit_orsim}. Given that the CIC statistics is indeed evaluating the PDF of galaxy number counts in cubic cells, we expect it to have a strong constraining power on the PDF of satellites within a halo.

The CIC estimator we consider, $n_{\rm CIC} \left(N_{\rm gal}\right)$, is obtained following a similar approach as \citep{Yang_2011}: the simulation box is divided in cubic cells of side $L_{\rm cell} = 5$ $\emph{h}^{-1}\rm{Mpc}$, then we count how many galaxies have each cell and finally we find out how many cells have $N_{\rm gal}$ galaxies. Finally, we normalize over the number of cells:

\begin{equation}
\label{eqn: CIC}
    n_{\rm CIC} \left(N_{\rm gal}\right) = N_{\rm cells} \left(N_{\rm gal}\right) \cdot \left(\frac{L_{\rm cell}}{L_{\rm box}}\right)^3
\end{equation}

$n_{\rm CIC} (N_{\rm sat})$ is estimated using 100 realizations of galaxy catalogues computed from \unit simulations, with $f_{\rm sat} = 0.3$ and $n = n_{\rm eBOSS}$. We point out that Counts-in-Cells, unlike two-point statistics, depends on the number density of the galaxy catalogue taken into account. If the number density is higher, we will have more cells filled with a higher number of galaxies, and less cells with a lower number of galaxies. 

\begin{figure}
    \centering
    \includegraphics[width=0.45\textwidth]{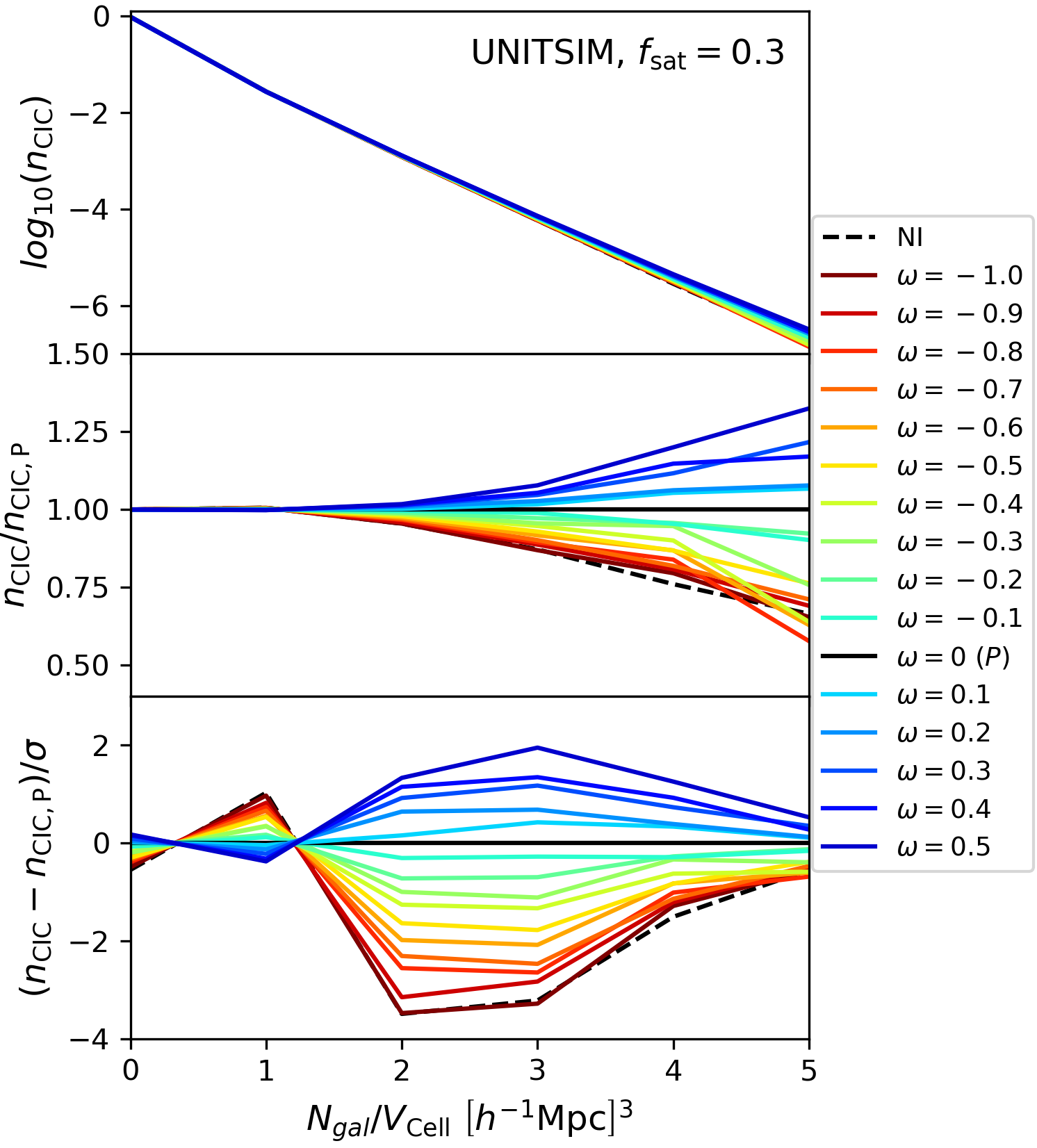}
    \caption{Count-in-Cells calculated from \unit simulations considering a fraction of satellites $f_{\rm sat} = 0.3$. We consider the negative binomial distribution ($\omega > 0$), Poisson distribution ($\omega = 0$), $B_{\rm sub-P}$ ($-1 \leq \omega < 0$) and Nearest Integer distribution.  {\it Top}: Count-in-Cells estimator $n_{\rm CIC} \left(N_{\rm gal}\right)$.  {\it Middle}: ratio between all $n_{\rm CIC}$ and Poisson $n_{\rm CIC,P}$ $(\omega = 0)$.  {\it Bottom}: difference between $n_{\rm CIC}$ and Poisson $n_{\rm CIC,P}$ divided by the Jackknife error $\sigma$.}
    \label{fig:CIC_all_PDF}
\end{figure}

In \autoref{fig:CIC_all_PDF} we can see how counts in cells vary with respect to $\omega$. We remind that $\omega$ is related to the variance of the distribution (See \autoref{eqn:variance_omega}).

In the upper panel we see the general trend of $n_{\rm CIC} (N_{\rm gal})$: fewer cubic cells in the box are expected to contain more galaxies. This decay is roughly exponential. 

The middle panel of \autoref{fig:CIC_all_PDF} shows better the variation on the number of cells depending on the value of $\omega$. All lines are divided by Poisson $n_{\rm CIC,P}$ ($\omega = 0$). For $N_{\rm gal}/V_{\rm cell} \geq 2$ we start to see that for $\omega > 0$, more number of cells are filled with higher number of galaxies. The trend is opposite for $\omega < 0$. 

In the lower panel we can see the potential of CIC constraining $\omega$. The highest constraining is achieved considering cells with 2 or 3 galaxies inside.  We can see also that noise increases for cells filled with higher number of galaxies.

\subsection{Proof of concept: using CIC to reproduce a model catalogue of galaxies}
\label{sec:CIC_and_clustering}

In this section we compare the constraining power of Count-in-Cells and two-point statistics, using \unit simulation. We use as our reference catalogue a galaxy catalogue with $(\omega , f_{\rm sat})= (-0.8, 0.4)$. We consider the same parameter space as in \autoref{sec: fitting_data}. For two-point statistics, the galaxy catalogues constructed have a number density $n_{\rm gal} = 6n_{\rm eBOSS}$, as in \autoref{sec: fitting_data}. In the case of CIC, since the results depend on the number density, we have constructed 10 realizations with number density $n_{\rm gal} = n_{\rm eBOSS}$ for each point in the parameter space to achieve a similar resolution.

We consider the same two-point statistics as before: ($w_{\rm p} (r_{\rm p})$, $\xi_0(s)$ and $\xi_2(s)$), but now we compare our galaxy catalogues to the reference galaxy catalogue generated with $\omega=-0.8$, $f_{\rm sat}=0.4$, instead of eBOSS data. For the Count-in-Cells information, we use $n_{\rm CIC} (N_{\rm gal})$ evaluated at $N_{\rm gal}/V_{\rm cell}$ from 0 to 5, since beyond some catalogues have no cells with $N_{\rm gal}/V_{\rm cell} > 5$. Finally, we also join all statistics. 

\begin{figure}
    \centering
    \includegraphics[width=0.48\textwidth]{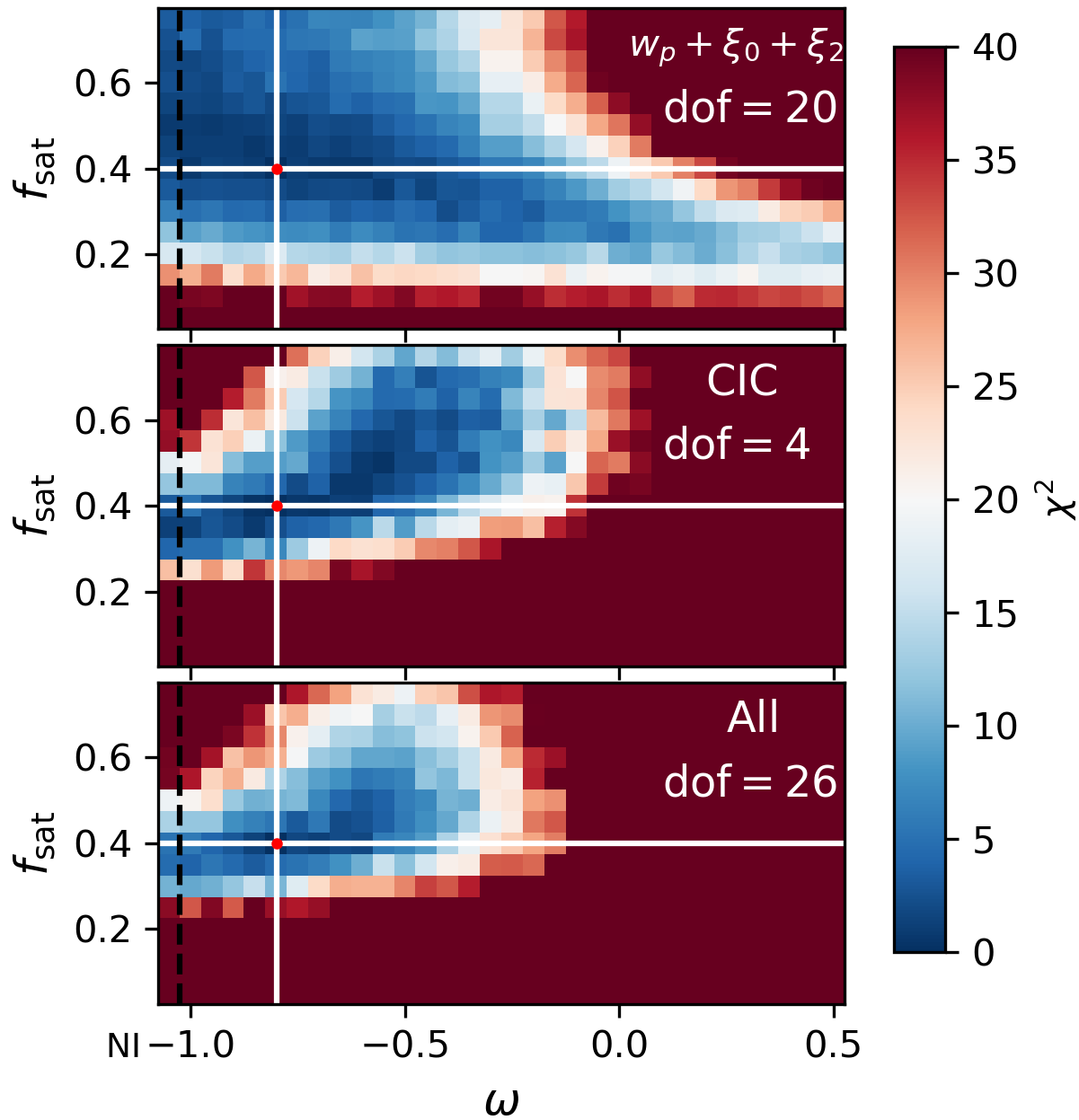}
    \caption{ {\it Top}: $\chi^2 (f_{\rm sat},\omega)$ obtained considering $w_{\rm p}$, $\xi_0$ and $\xi_2$ clustering measurements.  {\it Middle}: $\chi^2 (f_{\rm sat},\omega)$ obtained considering Count-in-Cells.  {\it Bottom}: $\chi^2$ joined from both contributions. Our reference galaxy catalogue has $(\omega , f_{\rm sat})= (-0.8, 0.4)$, with $\chi^2 = 0$.}
    \label{fig:xi2_clustering_CIC}
\end{figure}

In the top panel of \autoref{fig:xi2_clustering_CIC} we show the $\chi^2$ of the combined projected correlation function, monopole and quadrupole. As it can be seen, $f_{\rm sat}$ and $\omega$ are partially degenerated. That is, points far from each other in the parameter space may share similar $\chi^2$ near to the minimum.

In the middle panel we show the performance of the Count-in-Cells and in the bottom panel we sum CIC + 2PCFs. CIC turns to be promising when fitting the HOD parameters, since it has more constraining power on ($\omega,f_{\rm sat}$) than two-point functions. Also, it complements 2PCFs, providing a different orientation of the degeneracy contour, helping to break previous degeneracies between $f_{\rm sat}$ and $\omega$. 

We also expect other statistical methods based on counting galaxies in different volumes such as counts in cylinders and 2D k-Nearest Neighbours \citep{Yuan_2023} to be promising to break degeneracies in the parameter space. 

Nevertheless, despite CIC has potential as an statistic to describe the distribution of galaxies, more work needs to be done to properly account for data systematics such as the survey window, masks or redshift evolution of the number density $n(z)$ and linear bias $b(z)$, when fitting our galaxy catalogues to data surveys \citep{Salvador2019}. Moreover, the use of lightcone catalogues would also be more appropriate for this type of measurements.

\section{Summary and Conclusions}
\label{sec:conclusions}

We have generated catalogues with model galaxies using probability distribution functions (PDF) with a range of variances for deciding the number of satellite galaxies to be placed in a dark matter halo of a given mass. These catalogues are generated with HOD models with fixed number density and bias (\S~\ref{sec: HODmodel}). We have used dark matter haloes from the \unit and \orsim at $z = 0.8594$ and $0.865$, respectively.

In this work we have expanded the range of possible variances of the PDF for placing satellite galaxies in dark matter haloes with halo occupation distribution (HOD) models. In particular, we have made possible to have continuous sub-Poissonian variances ($\sigma^2<\lambda\equiv\langle N_{\rm sat}\rangle)$, introducing an extension to the binomial distribution, $B_{\rm ext} (N_{\rm sat}; \lambda,\omega)=f_{\rm q}B (N_{\rm sat};p,q)$ (\S~\ref{sec: PDFextensions}). $f_q$ is computed such that the extended binomial matches the moments of the binomial distribution (the calculations are detailed in appendix \ref{ap: f_calculation}).

We parameterize the PDF variance with $\omega_{\rm \sigma}$, $\sigma^2 = \lambda (1+\omega_\sigma \lambda)$ (\autoref{eqn:variance_omega}). This parameter quantifies the deviation of the variance with respect to one corresponding to a Poisson distribution, $\sigma^2 = \lambda$. The binomial distribution, $B (N_{\rm sat};p,q)$, can have discrete values for the variance $-1 \leq \omega_{\rm \sigma} = -\frac{1}{q = 1,2,...} < 0$ and the extended binomial distribution covers it continuously $-1 \leq \omega_{\rm \sigma} < 0$.

We find mathematical limitations for the  binomial (and also extended binomial) distribution functions. We introduce $\omega$ as the input parameter of $\omega_{\rm \sigma}$ in our code by the user. Then, the region of the parameter space delimited by ($\omega < -1/\lambda,\lambda > 1$), this PDF has negative probabilities. To correct this issue, the HOD code uses $\omega_{\rm \sigma} = -1/(\rm{trunc}(\lambda)+1)$ instead of $\omega$ (See \autoref{eqn: serrated}). This change allows to maximally explore sub-Poissonian variances with a same parameter $\omega$ for the entire catalogue without obtaining unphysical results. As the mean number of satellites in a halo typically increases with halo mass, $\lambda (M)$, this correction only is needed for high-mass haloes, which are a minority.

Negative probabilities can also arise for $B_{\rm ext}$ when considering a non-trivial subset of the following parameter space ($\lambda > 1,0>\omega > -\frac{1}{3}$). This is due to $f_q$, the factor that allows the variance of $B_{\rm ext}$ to be continuous, being negative. 
For the model catalogues we produce, this problem only affects $6$ per million haloes with an error in the mean number of satellites greater than $0.5 $ per cent, and 17 per million haloes are affected with an error in $\sigma$ greater than $0.5 $ per cent (See \autoref{sec: par_space_galaxy_catalogues}).

For all the PDFs for the number of satellite galaxies, we have mitigated a computational limitation that happens when using $\Gamma (x)$ for big enough arguments. This has been done by swapping ratios of $\Gamma (x)$ for products, as they are mathematically equivalent (\S~\ref{sec: gamma_limitation}).

Varying the PDF for satellite galaxies impacts the small scale clustering. PDFs with larger variances enhance the 1-halo projected two-point correlation function in redshift space in the scales $r_{\rm p} \lesssim 1$ $h^{-1} \rm{Mpc}$, beyond the statistical errorbars. The effect on the monopole and quadrupole of the two-point correlation function is negligible since we are using linear scales. 

Assuming different variances for the PDF of satellite galaxies also affects the Count-in-Cells (CIC). We have considered cubic cells of side $5 h^{-1} \rm{Mpc}$. This volume is small enough to see the PDF impact on CIC and big enough to make CIC calculations computationally feasible. The highest constraining power of CIC is found in the range $N_{\rm gal}/V_{\rm cell} = 2-3/\left[5 h^{-1} \rm{Mpc}\right]^3$ considering $n = n_{\rm eBOSS}$. 

We have applied our extension to find the best HOD model describing the clustering of eBOSS Emission-Line Galaxies (ELGs) at $z = 0.6 - 1.1$. For this exercise, only two parameters have been set free: $\omega$, which controls the variance of the PDF for satellite galaxies; and $f_{\rm sat}$, which controls the fraction of galaxies that are satellites (\S~\ref{sec: fitting_data}). We have fitted simultaneously the observed monopole, $\xi_0(s)$, quadrupole, $ \xi_2 (s)$ and the projected correlation function, $w_{\rm p} (r_{\rm p})$. For the error bars we have used jackknife covariance matrices (\S~\ref{sec:jackknife}). A sub-Poisson distribution together with a high fraction of satellites are preferred to reproduce the clustering of eBOSS ELGs. In particular, for \unit we find: $\omega = -0.65$, $f_{\rm sat} = 0.35$; and for \orsim: $\omega <1$ (minimal variance, corresponding to a nearest integer distribution), $f_{\rm sat} = 0.4$.

As an exercise to show the potential of CIC to constrain the PDF variance, and using a mock catalogue as a reference, we have included both the clustering and CIC to constrain the HOD model parameters. Again, we have only set free $\omega$ and $f_{\rm sat}$. CIC has more constraining power over the variance of the PDF for satellite galaxies. This estimator can reduce the degeneracies found between $\omega$ and $f_{\rm sat}$ when fitting only the clustering (\S~\ref{sec:CIC_and_clustering}).
However, more work needs to be done to fit mock catalogues to observed CIC, as several other aspects  need to be taken into account that were beyond the scope of this article. These include observational systematic errors, survey window function, the effect of density and bias evolution ($n(z)$, $b(z)$), etc.

In summary, we have developed a robust way to generate HOD galaxy catalogues with a range of variances for the PDF of satellite galaxies within dark matter haloes. We have shown that this variance has a large impact on one and two point summary statistics. In particular, CIC is a promising statistic to constrain the distribution of galaxies within dark matter haloes. Likewise, the PDF of satelllites needs to be understood well, if the CIC statistics is to be used to constrain cosmology.

Our extension to the HOD modelling and the proposal to use the observed CIC in the future give us the opportunity to gain insight of the galaxy-halo connection with existing and upcoming data. Reciprocally, it is fundamental to understand the impact of galaxy-halo connection ingredients on different galaxy clustering statistics, since some of these ingredients could be degenerated with cosmological parameters.
These topics are of great relevance now that we have entered the stage-IV of cosmological surveys with the data acquisition by the Dark Energy Spectroscopic Instrument (DESI) and Euclid.

\section{Data availability}

The data that support the findings of this study are available on request.

\section*{Acknowledgements}
We thank Andrew Hearin, Sihan Yuan and Antonela Taverna for useful discussions.

BVG, SA, VGP have been or are supported by the Atracci\'{o}n de Talento Contract no. 2019-T1/TIC-12702 granted by the Comunidad de Madrid in Spain. 
This work has also been supported by Ministerio de Ciencia e Innovaci\'{o}n (MICINN) under the following research grants: PID2021-122603NB-C21 (BVG, VGP, GY), PID2021-123012NB-C41 (main support for SA),  PID2021-123012NB-C43 (BVG) and PGC2018-094773-B-C32 (BVG, SA).
IFAE is partially funded by the CERCA program of the Generalitat de Catalunya.

The UNIT simulations have been run  in the MareNostrum
Supercomputer, hosted by the Barcelona Supercomputing Center,
Spain, under the PRACE project number 2016163937




\bibliographystyle{mnras}
\bibliography{HOD+ELG} 




\appendix

\section{Calculation of \lowercase{$ f_{\rm q} \left(\uppercase{N}_{\rm sat};\lambda,\omega \right)$}}
\label{ap: f_calculation}

In this appendix we detail the calculations to obtain an expression for $f_{\rm q} (N_{\rm sat};\lambda,\omega)$ (hereafter $f_{\rm q}(N_{\rm sat})$) and hence, the expression of the extended binomial distribution $B_{\rm ext}$ (See \autoref{eqn:extended_binomial}).

\subsection{Mapping between $q$ and $\omega$}
\label{ap: relation_q_omega}
For each $q$ value, $f_q (N_{\rm sat})$ is motivated for a particular subset of the continuous range in $\omega$: $q \equiv \rm{ceil}(-1/\omega)$  (See \autoref{sec:beyond_binomial} ). Then, we have that for $q = 1$, $\omega = -1$; for $q = 2$, $-\frac{1}{2} \geq \omega > -1$; for $q = 3$, $-\frac{1}{3} \geq \omega > -\frac{1}{2}$ and so on.

We motivate this mapping between $q$ and $\omega$. We have that if $q = 2$, the binomial distribution is defined for $N_{\rm sat} = 0,1,2$. As it will be explained later, three possible values of $N_{\rm sat}$ make possible to choose an expression of the variance as a function of $\omega$, such as \autoref{eqn:variance_omega}. In principle, we have that $\omega$ can take any value between -1 and 0, filling the whole range defining only one function: $f_{q=2}(N_{\rm sat})$. Nevertheless, $B_{\rm ext}$ will be defined always for $N_{\rm sat} = 0,1,2$, and in general, $B_{\rm ext}$ will not share the expressions of the higher order moments with the binomial distribution. Since we require $B_{\rm ext}$ to resemble as possible to the original binomial distribution, we want $q$ to be a natural number and same expressions for higher order moments. 

Then, if we consider $q = 3$, the expression of $f_{q=3} (N_{\rm sat}=3)$ becomes negative for $\omega < -\frac{1}{2}$ (See \autoref{eqn:f3_3}). In general, for $q = k$, the expression of $f_{q=k}(N_{\rm sat})$ becomes negative for $\omega < -\frac{1}{k-1}$. Finally, taking also into account that for the binomial distribution $q = -1/\omega$, the expression $q \equiv \rm{ceil}(-1/\omega)$ comes naturally to describe the mapping between both parameters.

\subsection{Computing \lowercase{$ f_{\rm q} \left(\uppercase{N}_{\rm sat};\lambda,\omega \right)$} for each $q$}

To determine $f_{\rm q} (N_{\rm sat};\lambda,\omega)$ we need to solve a different number of equations depending on the value of $q$. Then, $f_{\rm q} (N_{\rm sat};\lambda,\omega)$ has to be determined separately for each $q$.

We start at $q = 1$, which is the trivial case. As we know, this case applies only if the input parameter in the code is $\omega = -1$ . The binomial distribution is defined in the range $N_{\rm sat} = 0,1$, since $N_{\rm sat} \leq q$. In this particular case, to determine the expression of $f_{\rm q=1}(N_{\rm sat})$ we require only two equations, one for the normalization:

\begin{equation}
\label{eqn: norm1}
\begin{aligned}
    1 & = \sum_{N_{\rm sat}=0}^{q=1} B(N_{\rm sat}) f_1(N_{\rm sat}) = \left(1-\lambda\right) \cdot f_1\left(0\right) + \lambda \cdot f_1\left(1\right) \, ,
\end{aligned}
\end{equation}
and another one for the mean of the resulting distribution,

\begin{equation}
\label{eqn: mean1}
\begin{aligned}
    \lambda & = \sum_{N_{\rm sat}=0}^{q=1} N_{\rm sat} B(N_{\rm sat}) f_1(N_{\rm sat}) = \lambda \cdot f_1\left(1\right) \, ,
\end{aligned}
\end{equation}
in which $B(N_{\rm sat})$ is the binomial distribution defined in \autoref{sec: binomial}. We can see trivially that $f_1( N_{\rm sat}) = 1$ $\forall N_{\rm sat}$. That is, for $q=1$, no corrections are needed and $B_{\rm ext} ( N_{\rm sat})= B ( N_{\rm sat})$. As we have seen, we only have fixed the normalization and the mean of the resulting distribution, $B_{\rm ext}$. Higher order moment expressions cannot be chosen. In particular, the only possible expression of the second central moment, the variance, is:

\begin{equation}
    \sigma^2 = \sum_{N_{\rm sat}=0}^{q=1} N_{\rm sat}^2 B(N_{\rm sat}) f_1(N_{\rm sat}) - \lambda^2 = \lambda \left(1-\lambda \right) \, .
\end{equation}

This expression of the variance turns out that it coincides with \autoref{eqn:variance_omega} for $\omega = -1$ (See also \autoref{eqn: std_binomial} for $q=1$). Also, the only possible expression of the third central moment, the skewness is:

\begin{equation}
    k_3  = \frac{ \sum_{N_{\rm sat}=0}^{q=1} N_{\rm sat}^3 B(N_{\rm sat}) f_1(N_{\rm sat}) -3\lambda \sigma^2 - \lambda^3 }{\sigma^3}= \frac{1-2\lambda}{\sigma} 
\end{equation}

which is also equal to the binomial expression of the skewness for $q = 1$ (See \autoref{eqn:skewness}). 

If $q = 2$, this correspond to $-1 < \omega \leq -0.5$. The binomial distribution can take values at $N_{\rm sat} = 0,1,2$. Since the extended binomial distribution is defined in the same range, we need now three equations to determine the new values of the probabilities at $N_{\rm sat} = 0,1,2$. We consider an equation for the correct normalization of the new probability distribution function \footnote{Although we substitute $q = 2$ in $f_q (N_{\rm sat}) \xrightarrow[]{} f_2 (N_{\rm sat})$, we do not make yet this substitution in the actual expressions, since this helps to identify finally an expression of $f_q (N_{\rm sat})$. },

\begin{equation}
\label{eqn: norm1}
\begin{aligned}
    1 & = \sum_{N_{\rm sat}=0}^{q=2} B(N_{\rm sat}) f_2(N_{\rm sat}) =\\ 
      & =\left(1-\frac{\lambda}{q}\right)^2 \cdot f_2\left(0\right) + 2 \frac{\lambda}{q} \left(1-\frac{\lambda}{q}\right) \cdot f_2\left(1\right) + 
    \frac{\lambda^2}{q^2} \cdot f_2\left(2\right) \, ,
\end{aligned}
\end{equation}
an equation for the mean 

\begin{equation}
\label{eqn: mean1}
\begin{aligned}
    \lambda & = \sum_{N_{\rm sat}=0}^{q=2} N_{\rm sat} B(N_{\rm sat}) f_2(N_{\rm sat}) = \\
            & =2 \frac{\lambda}{q} \left(1-\frac{\lambda}{q}\right) \cdot f_2\left(1\right) + 2\frac{\lambda^2}{q^2} \cdot f_2\left(2\right) \, ,
\end{aligned}
\end{equation}
and another for the variance, in which we allow continuous values of $\omega$. 

\begin{equation}
\label{eqn: var1}
\begin{aligned}
    \sigma^2 & = \lambda \left(1+\omega \lambda\right) = \sum_{N_{\rm sat}=0}^{q=2} \left(N_{\rm sat}^2 B(N_{\rm sat}) f_2(N_{\rm sat})\right) - \lambda^2=  \\
             & = 2 \frac{\lambda}{q} \left(1-\frac{\lambda}{q}\right) \cdot f_2\left(1\right) + 4\frac{\lambda^2}{q^2} \cdot f_2\left(2\right) - \lambda^2 \, .
\end{aligned}
\end{equation}

We obtain the following exact solutions:

\begin{equation}
    f_2\left(2\right) = \frac{q^2}{2}\left(\omega + 1\right)
\end{equation}

\begin{equation}
    f_2\left(1\right) = \frac{q^2}{2} \frac{1-\lambda \left(\omega + 1\right)}{q \left(1-\frac{\lambda}{q}\right)}
\end{equation}

\begin{equation}
    f_2\left(0\right) = \frac{q^2}{2} \frac{1-\lambda + \frac{\lambda^2 \left(\omega+1\right)}{2}}{\frac{q^2}{2} \left(1-\frac{\lambda}{q}\right)^2}
\end{equation}

which can be generalised as:

\begin{equation}
\label{eq:f2}
    f_{q=2}\left(N_{\rm sat}\right) = \frac{2^2}{2} \frac{g_0 \left(N_{\rm sat}\right)+g_1 \left(N_{\rm sat}\right)+g_2 \left(N_{\rm sat}\right)}{\frac{2^{2-N_{\rm sat}}}{\left(2-N_{\rm sat}\right)!} \left(1-\frac{\lambda}{2}\right)^{2-N_{\rm sat}}}
\end{equation}

with

\begin{equation}
    g_0 \left(N_{\rm sat}\right) = \frac{\left(-1\right)^{-N_{\rm sat}}}{\left(-N_{\rm sat}\right)!}  \lambda^{-N_{\rm sat}} = 1 \hspace{2mm} \rm{for} \hspace{2mm}  N_{\rm sat} = 0 \hspace{2mm},
\end{equation}

\begin{equation}
    g_1 \left(N_{\rm sat}\right) = \frac{\left(-1\right)^{1-N_{\rm sat}}}{\left(1-N_{\rm sat}\right)!}  \lambda^{1-N_{\rm sat}} \hspace{2mm} \rm{for} \hspace{2mm} 1- N_{\rm sat} \geq 0 \hspace{2mm},
\end{equation}

and 
\begin{equation}
    g_2 \left(N_{\rm sat}\right) = \frac{\left(-1\right)^{2-N_{\rm sat}}}{\left(2-N_{\rm sat}\right)!} \left(\omega+1\right) \lambda^{2-N_{\rm sat}} \hspace{2mm} \rm{for} \hspace{2mm} 2- N_{\rm sat} \geq 0 \, .
\end{equation}

As in the $q = 1$ case, the skewness cannot be chosen for $q = 2$. Then, we can see if the expression of the binomial skewness given by \autoref{eqn:skewness} is also recovered for $q = 2$:

\begin{equation}
\begin{aligned}
    k_3  &= \frac{ \sum_{N_{\rm sat}=0}^{q=2} N_{\rm sat}^3 B(N_{\rm sat}) f_2(N_{\rm sat}) -3\lambda \sigma^2 - \lambda^3 }{\sigma^3} =\\
    & = \frac{1 - \lambda}{1 + \omega \lambda} \frac{1 + \lambda (1 + 3 \omega)}{\sigma} \, ,
\end{aligned}
\end{equation}
which is not equal to the expression of the binomial skewness \footnote{The same expression is recovered only for $\omega = -1$}. This issue supports the idea presented in \autoref{sec: binomial}, in which the binomial moment of order $k$ is only defined for $q \geq k$. Obviously, this is true for any probability distribution function, affecting $B_{\rm ext}$ as well.

For $q=3$, which corresponds to $-1/2 < \omega \leq -1/3$, the binomial distribution has the following range: $N_{\rm sat} \in \left[0,1,2,3\right]$. Thus, in this case we need to solve four equations: an equation for the normalization, 

\small

\begin{equation}
\label{eqn: norm2}
    1 = \left(1-\frac{\lambda}{q}\right)^3 \cdot f_3\left(0\right) + 3 \frac{\lambda}{q} \left(1-\frac{\lambda}{q}\right)^2 \cdot f_3\left(1\right) + 3 \frac{\lambda^2}{q^2} \left(1-\frac{\lambda}{q}\right) \cdot f_3\left(2\right) + \frac{\lambda^3}{q^3} \cdot f_3\left(3\right)
\end{equation}

\normalsize

for the mean 

\begin{equation}
\label{eqn: mean2}
    \lambda = 3 \frac{\lambda}{q} \left(1-\frac{\lambda}{q}\right)^2 \cdot f_3\left(1\right) + 2\cdot 3 \frac{\lambda^2}{q^2} \left(1-\frac{\lambda}{q}\right) \cdot f_3\left(2\right) + 3 \cdot \frac{\lambda^3}{q^3} \cdot f_3\left(3\right)
\end{equation}

the variance, 
\begin{equation}
\label{eqn: var2}
    \sigma^2 = 3 \frac{\lambda}{q} \left(1-\frac{\lambda}{q}\right)^2 \cdot f_3\left(1\right) + 2^2\cdot 3 \frac{\lambda^2}{q^2} \left(1-\frac{\lambda}{q}\right) \cdot f_3\left(2\right) + 3^2 \cdot \frac{\lambda^3}{q^3} \cdot f_3\left(3\right) - \lambda^2
\end{equation}

and a last equation, in which we 
make use of the expression of the third moment, the skewness, equating it to the binomial third moment: $k_3 = \frac{\left(1+2\lambda \omega\right)}{\sigma}$ (See \autoref{eqn:skewness}):

\begin{equation}
\begin{aligned}
\label{eqn: skew2}
    k_3 \sigma^3 &=3 \frac{\lambda}{q} \left(1-\frac{\lambda}{q}\right)^2 \cdot f_3\left(1\right) + 2^3\cdot 3 \frac{\lambda^2}{q^2} \left(1-\frac{\lambda}{q}\right) \cdot f_3\left(2\right) + \\
    & + 3^3 \cdot \frac{\lambda^3}{q^3} \cdot f_3\left(3\right) -3\lambda \sigma^2 - \lambda^3
\end{aligned}
\end{equation}

We find the correspondent solutions:

\begin{equation}
\label{eqn:f3_3}
    f_3\left(3\right) = \frac{q^3}{6} \left(\omega + 1\right)\left(2\omega + 1\right)
\end{equation}

\begin{equation}
    f_3\left(2\right) = \frac{q^3}{6}  \frac{\left(1+\omega\right)-\lambda \left(1+\omega\right)\left(2\omega + 1\right)}{q \left(1-\frac{\lambda}{q}\right)} 
\end{equation}

\begin{equation}
    f_3\left(1\right) = \frac{q^3}{6} \frac{1-\lambda \left(1+\omega\right) + \frac{1}{2} \lambda^2 \left(1+\omega\right)\left(2\omega + 1\right) }{\frac{q^2}{2} \left(1-\frac{\lambda}{q}\right)^2}
\end{equation}

\begin{equation}
    f_3\left(0\right) = \frac{q^3}{6} \frac{1-\lambda + \frac{1}{2}\lambda^2 \left(1+\omega\right) - \frac{1}{6} \lambda^3 \left(1+\omega\right)\left(2\omega + 1\right) }{\frac{q^3}{6} \left(1-\frac{\lambda}{q}\right)^3}
\end{equation}

which can be generalised again as:

\begin{equation}
\label{eq:f3}
    f_{q=3}\left(N_{\rm sat}\right) = \frac{3^3}{6} \frac{g_0(N_{\rm sat}) + g_1(N_{\rm sat}) + g_2 \left(N_{\rm sat}\right) +g_3 \left(N_{\rm sat}\right)}{\frac{3^{3-N_{\rm sat}}}{\left(3-N_{\rm sat}\right)!} \left(1-\frac{\lambda}{3}\right)^{3-N_{\rm sat}}}
\end{equation}

$g_3$ have the following expression:

\begin{equation}
    g_3 \left(N_{\rm sat}\right) = \frac{\left(-1\right)^{3-N_{\rm sat}}}{\left(3-N_{\rm sat}\right)!} \left(\omega+1\right) \left(2\omega + 1\right) \lambda^{3-N_{\rm sat}} \hspace{2mm} \rm{for} \hspace{2mm} 3- N_{\rm sat} \geq 0
\end{equation}

We have already obtained a solution for $f_{\rm q} (N_{\rm sat})$ when $q = 1$, $q = 2$ and $q = 3$. The idea is to generalise the result for all $q$. 

Taking into account the expressions of $f_{q=2}(N_{\rm sat})$ and $f_{q=3}(N_{\rm sat})$, we generalize for all $q$:

\begin{equation}
\label{eq:f}
    f_{\rm q}\left(N_{\rm sat}\right) = \frac{q^{N_{\rm sat}}\left(q-N_{\rm sat}\right)!}{q!} \frac{\sum_{n=N_{\rm sat}}^{q} g_n \left(N_{\rm sat}\right) }{ \left(1-\frac{\lambda}{q}\right)^{q-N_{\rm sat}}}
\end{equation}

with 

\begin{equation}
\label{eq:gn}
     g_n \left(N_{\rm sat}\right) = \frac{\left(-1\right)^{n-N_{\rm sat}}}{\left(n-N_{\rm sat}\right)!} \lambda^{n-N_{\rm sat}} \prod_{i=0}^{n-1} \left(i\omega+1\right) \hspace{2mm} \rm{for} \hspace{2mm} n- N_{\rm sat} \geq 0
\end{equation}

The extended binomial distribution $B_{\rm ext}$ is constructed as a product between the binomial distribution (See \autoref{sec: binomial}) and the additional exact function we have already computed (See \autoref{eq:f}): $B_{\rm ext} (N_{\rm sat}) = f_{\rm q} (N_{\rm sat}) \cdot B(N_{\rm sat})$. 

\subsection{Testing $B_{\rm ext}$ central moments}

The generalization for all $q$ expressed in \autoref{eq:f} is obtained identifying some factors in \autoref{eq:f2} and \autoref{eq:f3} to be proportional to $q$. Nevertheless, we need further calculations to already demonstrate that \autoref{eq:f} has the predicted mean, variance and skewness for any value of $q$. 

We can prove that $B_{\rm ext} (N_{\rm sat})$ has the same expressions of the mean, variance and skewness as the binomial distribution for any value of $q$, with $\omega$ as a continuous parameter, without the need of solving an increasing number of equations. In general, we can compute the expression of any moment of the distribution by taking derivatives of the moment generating function.

The moment generating function can be defined, for any discrete probability distribution function, as follows:

\begin{equation}
    M_{x}(t) = E(e^{tX}) = \sum_{x \in range(P(x))} e^{xt} P (x) 
\end{equation}
with $P(x)$ as our probability distribution function and $x = N_{\rm sat}$ in our work. For the extended binomial distribution we have that:

\begin{equation}
\label{eqn: mean_Mx}
\begin{aligned}
     M_{x}(t) & = \sum_{ x= 0}^q e^{xt} B_{\rm ext} (x) = \sum_{ x= 0}^q e^{xt} B (x) f_{\rm q} (x) = \\
            & = \sum_{ x= 0}^q e^{xt} \frac{q! p^{x} (1-p)^{q-x}}{x! (q-x)!} \frac{q^{x}(q-x)!}{q!} \frac{\sum_{n = x}^{q} g_n(x)}{(1-p)^{q-x}} = \\
            & = \sum_{ x= 0}^q \frac{e^{xt} (pq)^{x}}{x!} \left(\sum_{n = x}^{q} \frac{(-1)^{n-x}(pq)^{n-x}}{(n-x)!} \prod_{i=0}^{n-1} (i \omega + 1) \right)
\end{aligned}
\end{equation}

We show explicitly the computation of the estimation of the mean, the variance and the skewness. We start estimating the mean:

\begin{equation}
\label{eqn: mean_Ex}
\begin{aligned}
     E(X) = & \frac{d M_{x(t)}}{dt}|_{t = 0} = \\
     & = \sum_{ x= 0}^q \frac{\lambda^x \cdot xe^{x\cdot 0}}{x!} \left(\sum_{n = x}^{q} \frac{(-1)^{n-x}(pq)^{n-x}}{(n-x)!} \prod_{i=0}^{n-1} (i \omega + 1) \right) = \lambda 
\end{aligned}
\end{equation}

The extended binomial distribution has the following estimation of the mean:

\begin{equation}
    E(X) = \lambda
\end{equation}

The estimation of the variance is computed as follows:

\begin{equation}
\begin{aligned}
     E(X^2) = & \frac{d^2 M_{x(t)}}{dt^2}|_{t = 0} = \\
     & = \sum_{ x= 0}^q \frac{\lambda^x \cdot x^2}{x!} \left(\sum_{n = x}^{q} \frac{(-1)^{n-x}(pq)^{n-x}}{(n-x)!} \prod_{i=0}^{n-1} (i \omega + 1) \right) = \\
     & = \lambda^2 + \lambda (1 + \omega \lambda)  
\end{aligned}
\end{equation}

The variance recovered is:

\begin{equation}
    \sigma ^2 = E(X^2) - \left[E(X)\right]^2 = \lambda^2 + \lambda (1 + \omega \lambda) - \lambda^2 = \lambda (1 + \omega \lambda)
\end{equation}

Finally, we estimate the skewness:
\begin{equation}
\begin{aligned}
     E(X^3) = & \frac{d^3 M_{x(t)}}{dt^3}|_{t = 0} = \\
     & = \sum_{ x= 0}^q \frac{\lambda^x \cdot x^3}{x!} \left(\sum_{n = x}^{q} \frac{(-1)^{n-x}(pq)^{n-x}}{(n-x)!} \prod_{i=0}^{n-1} (i \omega + 1) \right) = \\
     & = \lambda + 3\lambda^2 \left(\omega + 1\right) + \lambda^3 \left(\omega + 1\right) \left(2 \omega + 1\right) 
\end{aligned}
\end{equation}

The skewness recovered is:

\begin{equation}
\begin{aligned}
    k_3 = & E\left[\left(\frac{X-\mu}{\sigma}\right)^3\right] = \frac{E(X^3) - 3\lambda \sigma^2 - \lambda^3}{\sigma^3} = \frac{1 + 2 \lambda \omega}{\sigma}
\end{aligned} 
\end{equation}

We showed that the mean, variance and skewness expressions are recovered for the extended binomial distribution for a general value of $q$.

\section{EZ mock covariance}
\label{ap:EZmocks}

We have obtained our results in this study relying on jackknife covariance measurements. Nevertheless, previous similar works such as \citet{Avila_2020} used a covariance calculated from EZ Mocks \citep{Zhao_2021}.

If we return to the $\chi^2$ expression we showed in \autoref{eqn: chi2data}, we now consider in this appendix the latter covariance, which is the result of the raw covariance matrix computation considering 1000 EZ Mocks and its posterior rescaling considering a best fit mock:

\begin{equation}
    C_{ij} = C^{\rm{EZ}}_{ij} \cdot \frac{y_{\rm{sim,i}} \cdot y_{\rm{sim,j}}}{y^{\rm{EZ}}_{i} \cdot y^{\rm{EZ}}_{j}}
\end{equation}

in which $C^{\rm{EZ}}$ is the raw covariance matrix of EZ mocks and, as before, $y = (w_{\rm p},\xi_0,\xi_2)$ and $\theta = \left(f_{\rm sat},\omega\right)$.
This rescaling is explained in more detail in \citet{Avila_2020}.

We consider the parameter space and the $\chi^2$ formalism described in \autoref{sec: fitting_data} with this new prospect of the covariance. 

\begin{figure}
    \centering
    \includegraphics[width=0.48\textwidth]{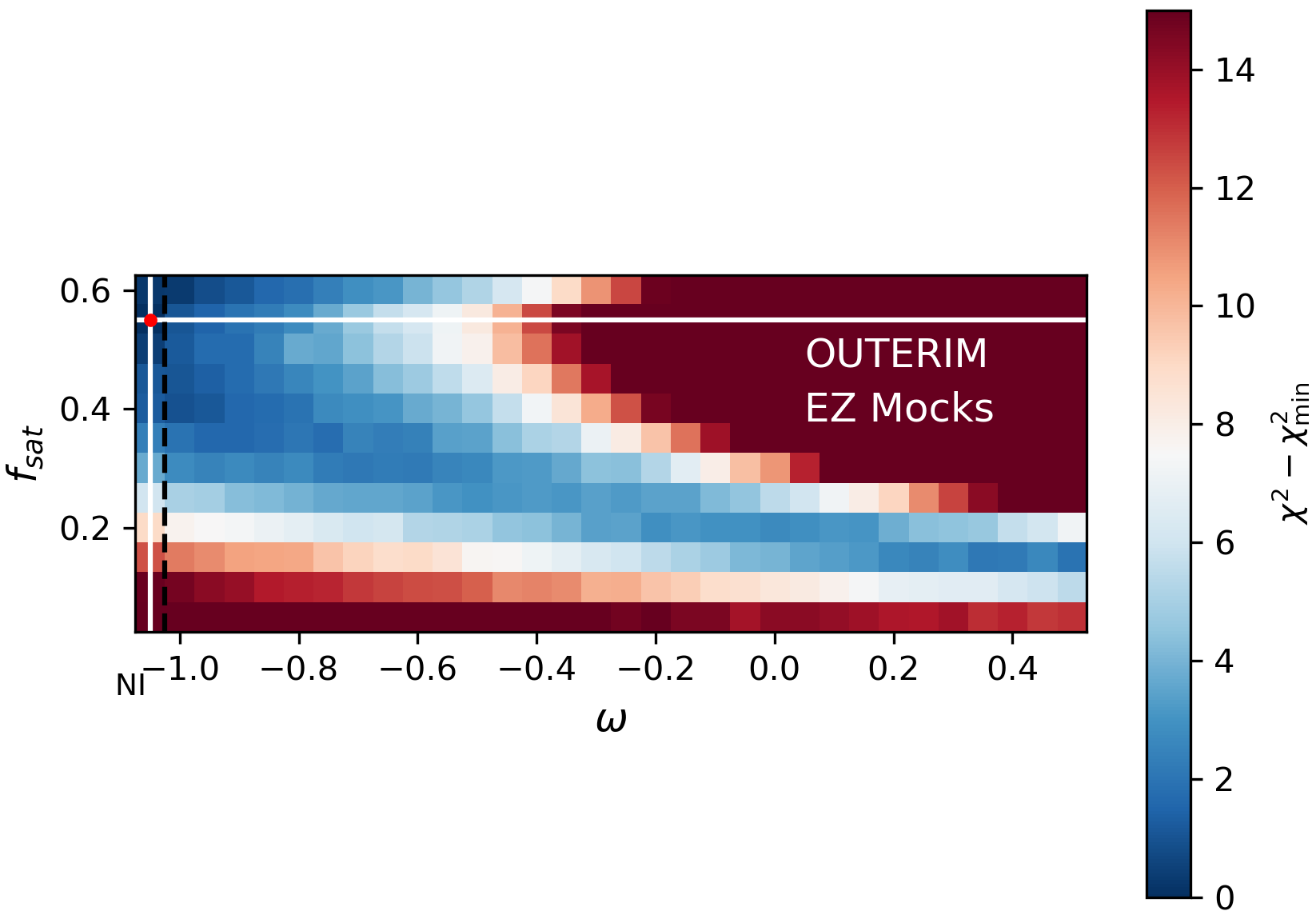}
    \caption{ Comparison between eBOSS ELG data and galaxy catalogues obtained form \orsim simulation with $\omega$ and $f_{\rm sat}$ as free parameters, using a covariance obtained from EZ mocks. The best fit to the data prefers a Nearest Integer distribution and a fraction of satellites  $f_{\rm sat}= 0.55$. \citet{Avila_2020} found a similar best fit using the same simulation, covariance and free parameters ($f_{\rm sat}$, $\omega$).}
    \label{fig:xi2_orsim_ezmock}
\end{figure}

\autoref{fig:xi2_orsim_ezmock} shows the results only considering the \orsim simulation. The best fit prefers a Nearest Integer distribution as in the lower panel of  \autoref{fig:xi2_unit_orsim}, but the fraction of satellites has increased from $f_{\rm sat} = 0.4$, that we had obtained with the jackknife covariance, to $f_{\rm sat} = 0.55$ using EZ mock covariance. On the other hand, \citet{Avila_2020} found very similar results, preferring a Nearest Integer distribution and $f_{\rm sat} = 0.56$.

We can see that in general, $\chi^2$ values are lower compared to the results shown in \autoref{fig:xi2_unit_orsim}. This effect arises 
because the value of the errors increase using EZ mock covariances when considering small scales (precisely, in the projected correlation function), 
compared to jackknife covariances. We have to take into account that EZ mocks are not expected to reproduce well enough the signal and variance of the small-scale regime, thus explaining the discrepancies we found comparing with the jackknife prescription.

Introducing this new covariance represents a modification of the minimum $\chi^2$ value and its location, although the overall contour look similar. We leave for future work when dealing with new data a more thorough validation and characterisation of the covariance. 

\bsp	
\label{lastpage}
\end{document}